\documentclass[letterpaper,twocolumn,10pt]{article}
\usepackage{usenix}

\usepackage{tikz}
\usepackage{times}
\usepackage{latexsym}
\usepackage{amsmath,amssymb,amsfonts}
\usepackage{graphicx}
\usepackage{textcomp}
\usepackage{xcolor}
\usepackage{url}
\usepackage{breakurl}
\usepackage{hyperref}
\usepackage{subcaption}
\usepackage{float}
\usepackage{booktabs}
\usepackage{algorithm}
\usepackage{algpseudocode}
\usepackage{float}
\usepackage{pifont}

\usepackage{filecontents}

\begin{document}

\date{}

\title{HillInfer: Efficient Long-Context LLM Inference on the Edge with Hierarchical KV Eviction using SmartSSD}

\author{He Sun$^{1}$, Shinan Liu$^2$, Li Li$^3$, Mingjun Xiao$^1$\\
  1 Department of Computer Science and Technology\&
  Suzhou Institute for Advanced Study\&\\
  State Key Laboratory of Cognitive Intelligence, University of Science and Technology of China\\
  2 Department of Data and Systems Engineering, University of Hong Kong\\
  3 State Key Laboratory for Internet of Things for Smart City, University of Macau\\
  Email: hesun@mail.ustc.edu.cn,\ shinan6@hku.hk,\ llili@um.edu.mo,\ xiaomj@ustc.edu.cn
}

\maketitle

\begin{abstract}
Deploying Large Language Models (LLMs) on memory-constrained AI Personal Computers (AIPCs) enables low-latency, privacy-preserving inference, but long-context generation is fundamentally bottlenecked by the linearly growing Key-Value (KV) cache. While dynamic KV eviction mitigates this memory wall, existing offloading strategies either trigger crippling PCIe I/O bottlenecks on standard SSDs or suffer from FPGA resource exhaustion by forcing compute-intensive exact attention on a single, weak Computational Storage Drive (CSD). In this paper, we propose HillInfer, a CSD-assisted KV eviction framework that introduces a paradigm shift: offloading strictly lightweight token importance evaluation to a single CSD (e.g., SmartSSD) on AIPCs. To fully capitalize on this lightweight offloading strategy, HillInfer orchestrates a Hierarchical KV Cache Manager (HKM) that leverages temporal locality and dynamic token hit rates to physically partition cache pools, thereby eliminating cross-device I/O thrashing. Additionally, we design an Adaptive Prefetch-based Pipeline (APP) that adaptively balances the evaluation workload between the host CPU and the SmartSSD, effectively masking the heterogeneous straggler effect. Finally, we introduce a CSD-based Evaluation Configuration (CEC) to enable resource-efficient near-data processing on the FPGA. Extensive experiments on a commodity AIPC demonstrate that HillInfer achieves up to an 8.56$\times$ speedup over state-of-the-art baselines, delivering low-latency, I/O-efficient long-context inference without sacrificing model accuracy.
\end{abstract}

\section{Introduction}
Large Language Models (LLMs) \cite{zhao2023survey,chang2024survey,minaee2024large,naveed2025comprehensive,li2024survey,friha2024llm} are transforming artificial intelligence by demonstrating exceptional capabilities in understanding natural language and supporting a wide range of language-centric downstream tasks \cite{isik2024scaling,wei2021pretrained,oyelade2025survey}. In practice, many of these tasks rely on prompts that inherently involve sensitive information, such as personal medical records, proprietary creative content, and confidential business data. This privacy concern has driven a growing trend toward deploying LLMs on AI Personal Computers (AIPCs), pushing for locally executed, secure LLM inference \cite{zheng2026solidattention,cai2024edge,yu2024edge,lu2024small}.

However, the limited memory capacity and computational resources of commodity AIPCs pose fundamental challenges to efficient LLM deployment \cite{tian2025clone,yin2024llm}. Even worse, the growing complexity of downstream tasks (e.g., code understanding, autonomous agents, and legal document analysis) increasingly requires LLMs to process massive input contexts \cite{liu2024cachegen,sun2025breaking,lin2024infinite}. As a result, beyond the static memory occupied by model parameters, the KV cache \cite{li2024survey}, whose footprint grows linearly with context length, has emerged as the primary memory bottleneck for long-context inference on AIPCs \cite{sun2025breaking,zhang2025pqcache,hooper2025squeezed}. For example, inferring the Qwen-7B \cite{bai2023qwen} model with a 4K context and a batch size of 8 consumes approximately 30 GB of memory. This far surpasses the 24 GB VRAM limit of high-end commodity GPUs like the NVIDIA RTX 4090, inevitably triggering out-of-memory (OOM) failures.

\noindent\textbf{Limitations of Prior Arts.}
Inspired by the transformer architecture and linguistic characteristics \cite{vaswani2017attention,beltagy2020longformer,levy2014linguistic}, long-context inputs exhibit a high degree of sparsity, where only a small subset of tokens (typically less than 20\% \cite{sun2025breaking,lee2024infinigen,tang2024quest}) plays a critical role in attention computations. Existing H2O-like \cite{zhang2023h2o} approaches leverage this observation by estimating token importance, selectively retaining the KV data of critical tokens on the GPU, and evicting less important ones during each autoregressive decoding step \cite{lee2024infinigen,zhang2023h2o,zhao2024alisa,gao2024cost,tang2024quest,liu2023scissorhands,juravskyhydragen,ye2024chunkattention,zheng2024sglang}. However, these methods are primarily designed for industrial-grade GPUs with massive memory capacities, restricting KV cache offloading strictly between the GPU and the host CPU. On memory-constrained AIPCs, the KV cache generated by long-context inference rapidly exceeds the combined capacity of the GPU and CPU, making offloading to external storage inevitable. Recent studies \cite{sun2025breaking,sheng2023flexgen,chen2025impress} explore standard NVMe SSDs to assist KV cache management. However, because dynamic eviction requires evaluating token importance at every decoding step, massive KV data must be repeatedly transferred between the SSD and the host. This data movement across the limited SSD bandwidth scales poorly with context length, eventually eclipsing the computation time and becoming the dominant inference bottleneck. To mitigate these severe I/O bottlenecks, recent Computational Storage Device (CSD) based frameworks, such as InstInfer \cite{pan2025instattention} and HILOS \cite{jang2025inf}, attempt to perform near-data attention. Unfortunately, calculating exact attention weights requires complex Softmax operations that quickly exhaust the strict compute and logic budgets of commodity FPGAs. To compensate, these approaches are forced to rely on heavily specialized custom hardware or multi-device scale-out architectures tailored exclusively for offline batched inference. Consequently, they inherently mismatch with the dynamic, step-by-step KV eviction paradigm, failing to support low-latency, long-context inference on a single commodity AIPC. (See all details of exploring existing techniques in Sec. \ref{Sec:EET})

\noindent\textbf{A Paradigm Shift: In-Storage Token Evaluation.} 
To break this impasse, we must rethink the role of external storage in LLM serving. Rather than forcing the CSD to execute resource-heavy exact attention computations, we observe a paradigm-shifting opportunity: offloading only the lightweight token importance evaluation to the storage layer. By integrating a commercial SmartSSD \cite{samsungSmartSSD} and executing this scoring phase locally on the onboard FPGA, we can exploit Near-Data Processing (NDP) to filter out unimportant tokens directly at the source. This fundamentally eliminates the redundant host-SSD data movement that plagues standard SSDs, while completely obviating the need for expensive multi-drive CSD deployments seen in prior approaches..

\noindent\textbf{Challenges.}
However, realizing this CSD-assisted KV eviction paradigm introduces severe system-level challenges. \textit{First}, token importance is inherently query-dependent \cite{sun2025breaking,lee2024infinigen,tang2024quest}, meaning a token's importance score fluctuates dramatically across decoding steps. Under a naive in-storage eviction strategy, this dynamic behavior triggers frequent ping-pong data movement (I/O thrashing) between the host CPU and the CSD, offsetting the benefits of near-data computation. \textit{Second}, coordinating this heterogeneous AIPC environment is non-trivial due to a drastic performance mismatch. The multi-core CPU and the CSD's embedded FPGA possess entirely different processing throughputs. If not carefully orchestrated, the faster processor will finish early and stall, leading to a severe straggler effect that prevents the system from effectively hiding the KV evaluation and transmission latency behind the GPU's computation.

\noindent\textbf{Our Contributions.}
In this paper, we propose \textbf{HillInfer}, an efficient, CSD-assisted KV eviction framework for long-context LLM inference in memory-constrained AIPCs. HillInfer orchestrates a hierarchical KV cache eviction strategy and an adaptive prefetch-based pipeline using a single commodity SmartSSD to systematically tackle the I/O thrashing and performance mismatch. Specifically, to overcome the first challenge, we design a Hierarchical KV Cache Manager (HKM) that physically partitions hot and cold KV caches between the CPU and the SmartSSD. By jointly optimizing for temporal locality and dynamic token hit rates, HKM effectively suppresses cross-device I/O thrashing (ping-pong data movement). To address the second challenge, we introduce an Adaptive Prefetch-based Pipeline (APP). APP dynamically balances the distributed evaluation workload between the CPU and the SmartSSD, achieving optimal I/O-compute overlap and seamlessly hiding the heterogeneous evaluation latency. Furthermore, to physically realize this NDP without exhausting FPGA resources, we introduce a CSD-based Evaluation Configuration (CEC). By completely stripping away resource-heavy exact attention computations in favor of hardware-efficient, asymmetric-precision streaming execution, CEC makes lightweight in-storage evaluation feasible on a single device. The main contributions of this paper are summarized as follows:

\begin{itemize}
    \item We propose HillInfer, to the best of our knowledge, the first long-context LLM inference framework on memory-constrained AIPCs that enables hierarchical, importance-aware KV cache eviction using a single commodity SmartSSD, effectively reducing end-to-end inference latency while preserving model accuracy.
    \item We design a Hierarchical KV Cache Manager featuring bidirectional cache pools to drastically reduce ping-pong I/O thrashing, coupled with a system-level Adaptive Prefetch-based Pipeline to eliminate the straggler effect and minimize end-to-end GPU idle time.
    \item We introduce a CSD-based Evaluation Configuration to exploit algorithmic simplification and asymmetric precision and implement HillInfer based on an offloading-based LLM inference framework and develop an HLS-based FPGA program on SmartSSD to support in-storage importance evaluation. 
    \item We conduct extensive experiments across diverse models and datasets. Evaluations demonstrate that HillInfer achieves up to an 8.56$\times$ speedup over state-of-the-art baselines without sacrificing generation quality.
\end{itemize}

\section{Background}
\subsection{LLM Inference}
\begin{figure*}[tbp]
    \centering
    \includegraphics[width=0.99\textwidth]{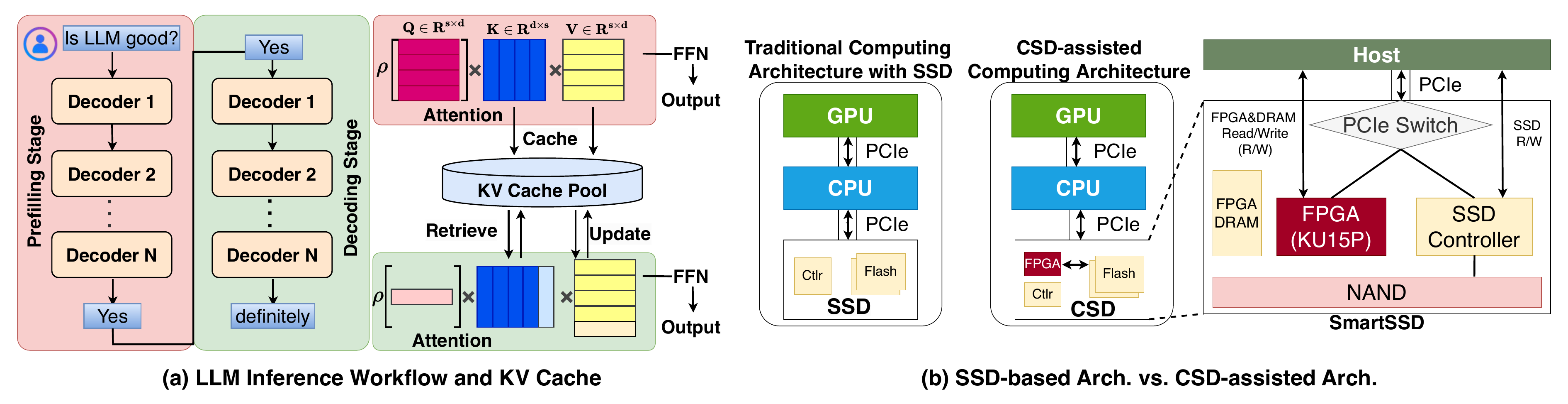}
    \caption{LLM and CSD Background: (a) An example of LLM Inference Workflow and illustration of KV Cache. (b) Traditional SSD-based Architecture. vs. CSD-assisted Architecture (e.g., SmartSSD).}
    \label{fig:CSD}
\end{figure*}
\textbf{LLM Inference Workflow.} Large Language Model (LLM) inference is executed in two distinct stages: \textit{prefilling} and \textit{decoding} \cite{zhao2023survey, minaee2024large}. 
During the prefilling stage, the tokenized input prompt is processed in a single forward pass to initialize the model's internal states. 
Subsequently, the decoding stage generates tokens autoregressively; each newly generated token is appended to the existing context and fed back into the model to predict the next token. 
This iterative process continues until an end-of-sequence (EOS) token is produced or a predefined maximum length is reached. 
In practice, input text is partitioned into units called tokens; for example, as illustrated in Figure \ref{fig:CSD} (a), the query ``Is LLM good?'' is tokenized into four discrete elements (``Is'', ``LLM'', ``good'', and ``?''). 
After processing these tokens through $N$ decoder layers during prefilling, the LLM produces an initial response token, such as ``Yes''. 
This token is then concatenated with the original prompt to serve as the context for generating the subsequent token, e.g., ``definitely''.

As shown in Figure \ref{fig:CSD} (a), modern LLM architectures are predominantly composed of Transformer decoder layers, each consisting of an attention mechanism and a Feed-Forward Network (FFN). 
Within each layer, hidden states are linearly projected into Query ($Q$), Key ($K$), and Value ($V$) tensors. 
The attention mechanism captures contextual dependencies by computing the weighted sum of values based on the similarity between queries and keys:
\begin{equation}\label{Eqs:attn}
    Attn(Q,K,V)=\rho\left(\frac{Q\cdot K^{T}}{\sqrt{d}}\right)\cdot V
\end{equation}
To enhance model capacity, advanced variants such as Multi-Head Attention, Multi-Query Attention (MQA) or Grouped-Query Attention (GQA), are employed to attend to multiple semantic subspaces in parallel.

\noindent\textbf{KV Cache Mechanism.} As defined in Equation \ref{Eqs:attn}, the self-attention operation inherently entails quadratic computational complexity relative to the input sequence length~\cite{beltagy2020longformer}. This quadratic growth, illustrated in Figure~\ref{fig:CSD} (a), stems from the requirement to calculate the full attention weight matrix across all tokens in every iteration~\cite{tang2024quest,lee2024infinigen}. Such overhead becomes a primary inhibitor for low-latency inference as the context length increases, leading to prohibitive latency~\cite{wu2024loongserve,sun2025breaking,hooper2025squeezed}.

To bypass this bottleneck, modern LLM engines utilize KV Cache to transform the decoding workload. Instead of re-evaluating the entire sequence, the system persists the Key ($K$) and Value ($V$) tensors of previous tokens in memory \cite{zheng2024sglang,kwon2023efficient}. During each subsequent decoding step, only the $Q, K,$ and $V$ tensors for the single newly generated token are computed. As shown in the data flow of Figure \ref{fig:CSD} (a), the current Query ($Q$) is then matched against the cumulative $K$ and $V$ tensors (the KV cache). This shift effectively reduces the computational complexity from quadratic to linear, enabling near-constant execution time per decoding step. 

\begin{figure}[tbp]
    \centering
    \begin{minipage}{0.23\textwidth}
        \centering
        \includegraphics[width=\textwidth]{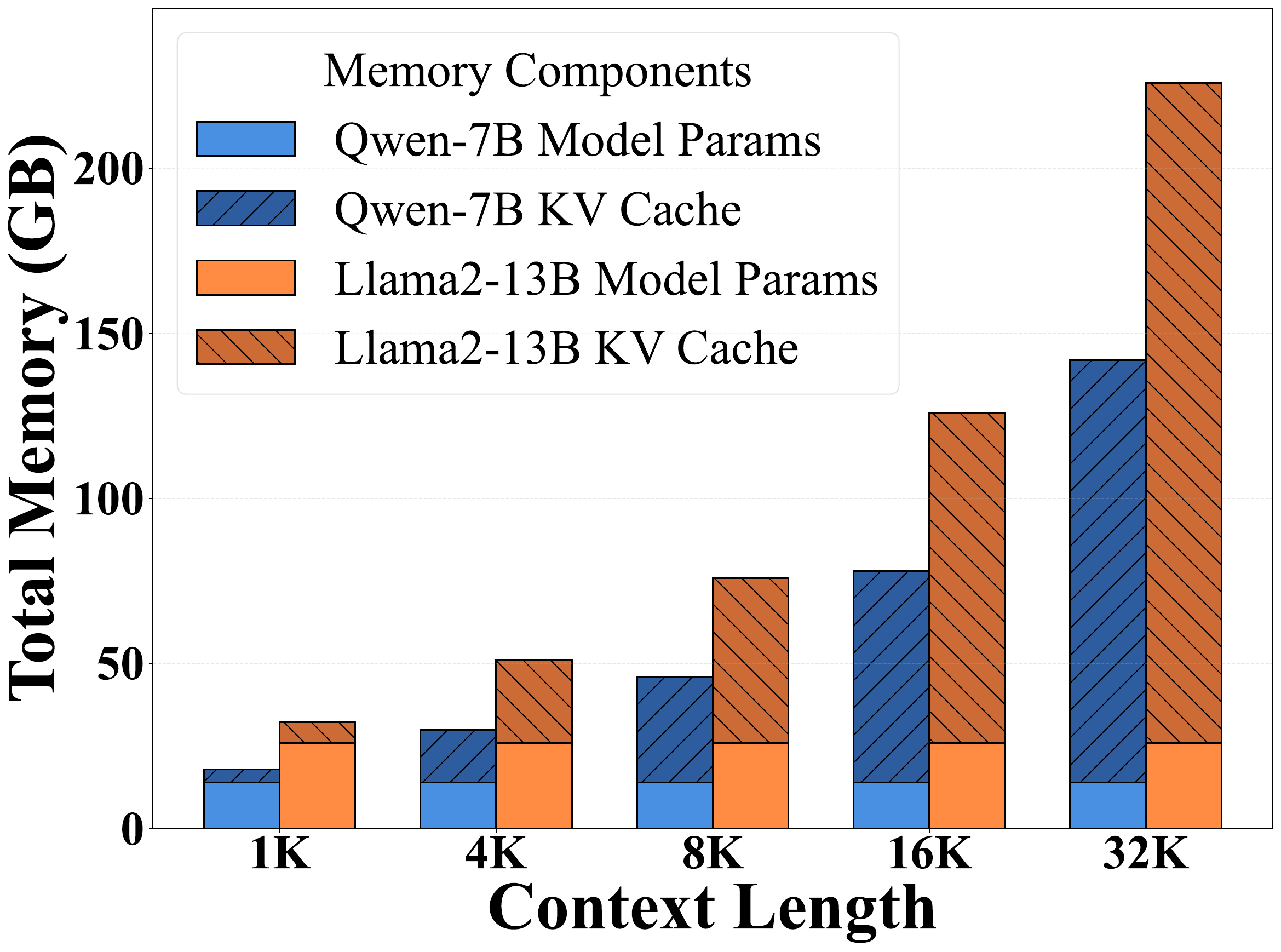}
        \subcaption{Memory usage with batch size = 8, varying context lengths.}
        \label{fig:context_length}
    \end{minipage}
    \hfill
    \begin{minipage}{0.23\textwidth}
        \centering
        \includegraphics[width=\textwidth]{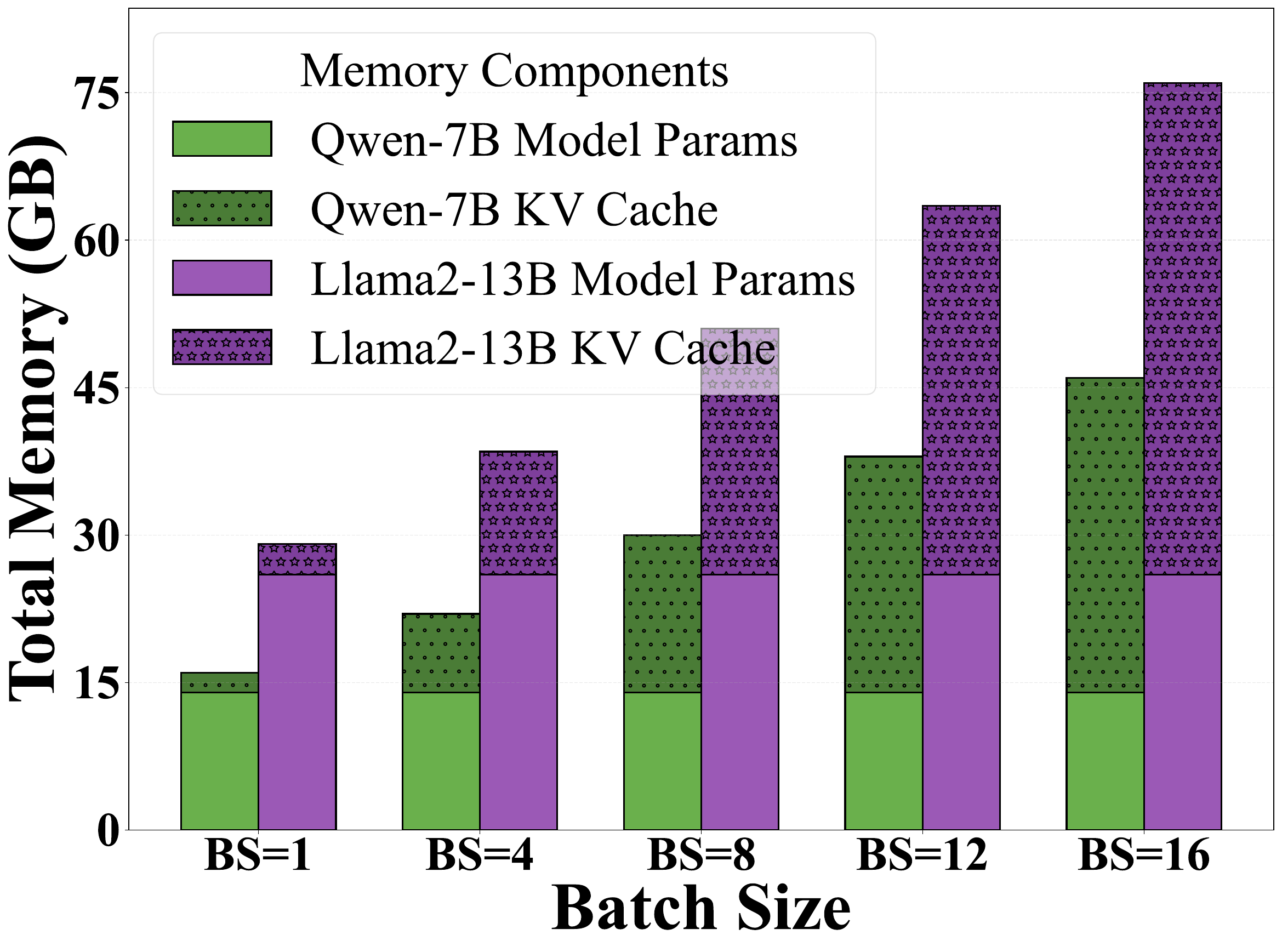}
        \subcaption{Memory usage with context length = 4K, varying batch sizes.}
        \label{fig:batch_size}
    \end{minipage}
    \caption{Memory usage analysis for Qwen-7B and Llama2-13B models with full KV cache. (a) shows how memory increases with context length at a fixed batch size of 8. (b) shows how memory scales with batch size at a fixed context length of 4K.}
    \label{fig:memory_analysis}
\end{figure}

\noindent\textbf{The Memory Wall in Long-Context Inference.} Despite the computational benefits of KV caching, it imposes a severe memory tax that grows proportionally with context length, the number of decoder layers, and the hidden dimension of the model. In memory-constrained environments, such as commodity PCs, this linear expansion creates a significant ``memory wall.'' For instance, as context length increases, the cumulative footprint of $K$ and $V$ tensors can quickly surpass the limited VRAM of entry-level GPUs and the available host DRAM \cite{zheng2026solidattention}. As illustrated in Figure \ref{fig:context_length}, the system experiences a fundamental bottleneck shift: while initial decoding steps may be compute-bound, long-context inference eventually becomes memory-bound. When the KV cache exceeds the aggregate capacity of volatile memory, the system must offload data to external SSDs to maintain stability \cite{sheng2023flexgen,sun2025breaking,chen2025impress,zheng2026solidattention}. However, traditional SSD-based offloading reintroduces prohibitive I/O latencies due to the frequent movement of massive KV data across the saturated PCIe bus (See details in Sec. \ref{Sec:EET}). This hardware-level bottleneck necessitates a paradigm shift toward processing data closer to the storage medium to eliminate redundant data transfers.

\subsection{Computational Storage}

A Computational Storage Device (CSD) \cite{samsungSmartSSD} enables Near-Data Processing (NDP) by embedding computing resources directly into the storage medium. As contrasted in Figure \ref{fig:CSD} (b), traditional SSD architectures are host-centric, forcing all data to traverse the external PCIe bus for computation, which creates severe bandwidth bottlenecks. Conversely, CSDs utilize an internal PCIe switch to orchestrate data flow locally, effectively minimizing external data movement.

Commercial implementations like the Samsung SmartSSD \cite{samsungSmartSSD} synergize an FPGA (e.g., Xilinx KU15P) and on-board DRAM alongside standard SSD components. Adhering to standard PCIe or U.2 form factors, it offers plug-and-play scalability for existing commodity PCs without hardware modifications. Crucially, while the host manages the device via standard I/O interfaces, the data path between the NAND flash and the FPGA is handled via internal Peer-to-Peer (P2P) transfers through the shared on-board DRAM buffer. This P2P architecture completely bypasses the host CPU and external PCIe bus, eliminating the data movement overhead of traditional SSD offloading and allowing computational storage capacity to scale linearly as additional devices are integrated.

\section{Motivations}
\subsection{Exploring Existing Techniques}\label{Sec:EET}

\textbf{Sparsity Opportunity: KV Cache Eviction.} In long-context scenarios, LLM attention mechanisms exhibit extreme sparsity, with typically less than 20\% of historical tokens significantly influencing the final output \cite{zhang2023h2o,sun2025breaking,tang2024quest}. To exploit this, modern inference engines employ a dynamic KV cache eviction workflow \cite{zhang2023h2o,feng2024ada,sun2025breaking,tang2024quest,zhang2025pqcache,ge2024model,xiaoefficient}. Specifically, during each step of the autoregressive decoding stage, the system computes an importance score for all historical tokens based on the current query. The engine then selectively retains only the highly-scoring top-$\alpha$ percent KV data in the active GPU cache, while evicting, offloading, or dropping the less critical ones into host memory or even disk \cite{sun2025breaking,chen2025impress,zheng2026solidattention}. By iteratively evaluating and filtering tokens at runtime, this importance-aware strategy significantly reduces the peak memory footprint required for long-context generation.

\noindent\textbf{The Limits of GPU-CPU Offloading.} A common approach to mitigate the KV cache burden is exploiting contextual sparsity to estimate token importance and selectively evict less critical KV entries \cite{zhang2023h2o,lee2024infinigen,zhao2024alisa,gao2024cost,tang2024quest}. However, these algorithms are primarily designed for data-center environments equipped with industrial-grade GPUs and massive host memory, relying heavily on two-level GPU-CPU offloading. When deployed on memory-constrained PCs, this architecture hits a hard capacity wall. As illustrated in Figure \ref{fig:memory_analysis}, increasing either the context length or the batch size causes the memory footprint of the KV cache to explode, quickly exceeding the combined capacity of both the GPU VRAM and the host DRAM (e.g., a 13B model with a 16K context easily exhausts the 84GB combined memory of an RTX 4090 AIPC). While aggressive eviction or memory pooling can delay OOM, they force the system to repeatedly recompute discarded KV states, introducing prohibitive computational overhead that negates the fundamental benefits of caching.

\begin{figure}[tbp]
    \centering
    \begin{minipage}{0.25\textwidth}
        \centering
        \includegraphics[width=\textwidth]{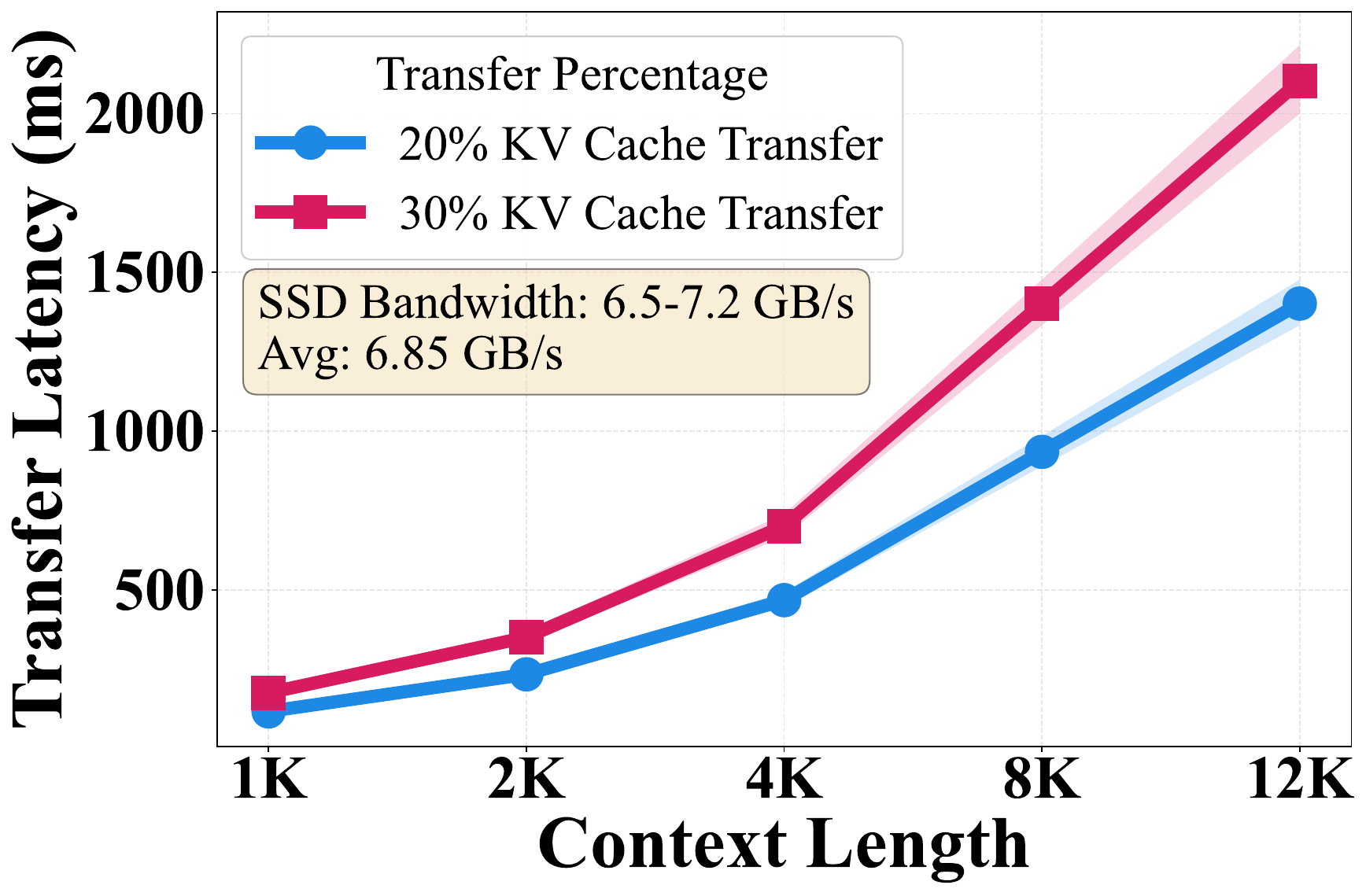}
        \subcaption{KV Cache Transfer Latency.}
        \label{fig:2-1}
    \end{minipage}
    \hfill
    \begin{minipage}{0.22\textwidth}
        \centering
        \includegraphics[width=\textwidth]{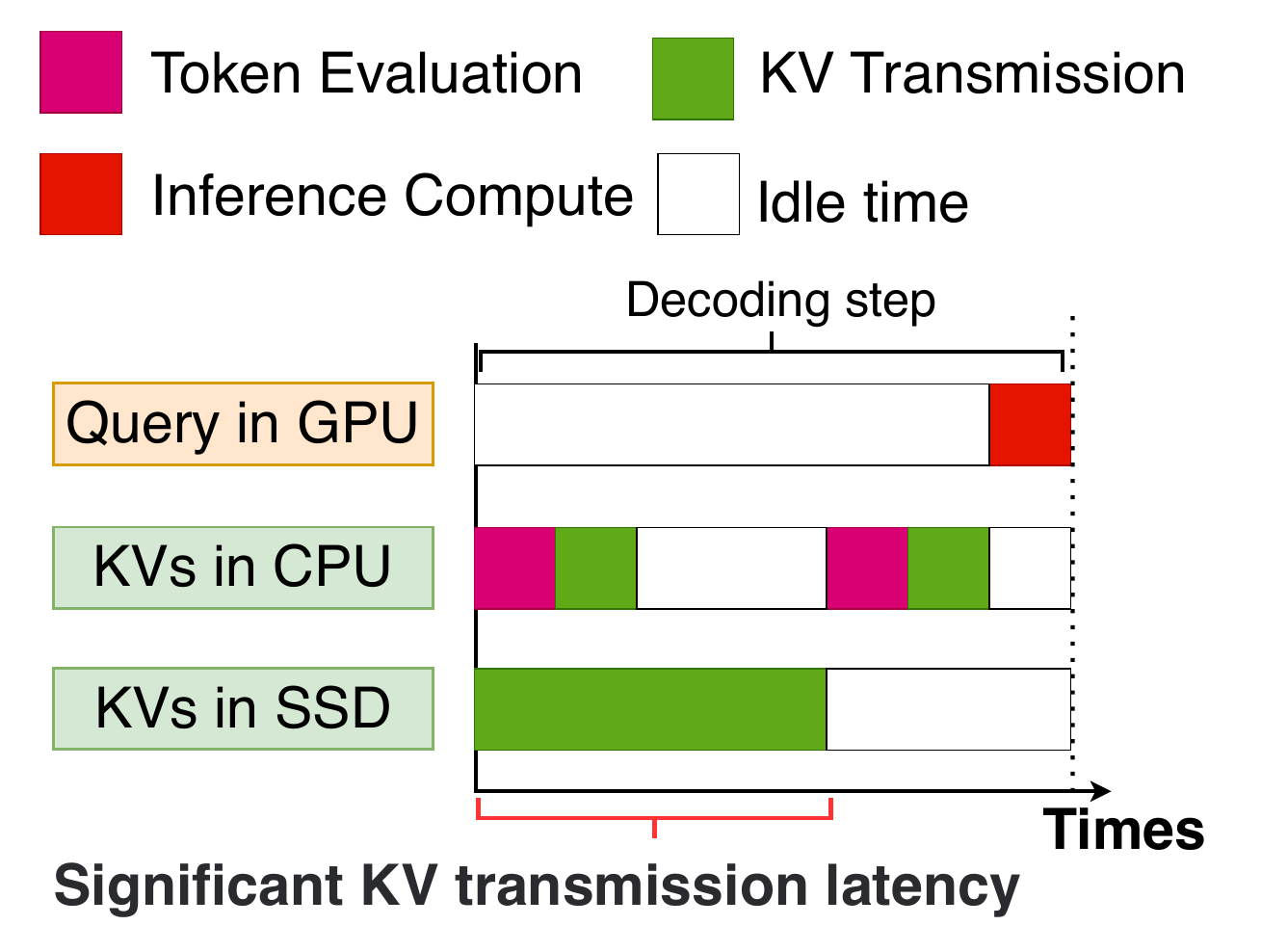}
        \subcaption{Latency Analysis.}
        \label{fig:2-2}
    \end{minipage}
    \caption{KV Cache Transfer Latency Analysis: (a) It shows the transfer time from SSD to GPU/CPU for 20\% (blue) and 30\% (red) of full KV cache across different context lengths for token evaluations at a batch size of 8 using Qwen-7B; (b) The illustration of the overhead of KV Cache transfer latency.} 
    \label{fig:kv_transfer_latency}
\end{figure}

\noindent\textbf{The I/O Bottleneck in SSD-Assisted Serving.} To bypass volatile memory limits, recent systems leverage external storage (e.g., NVMe SSDs) to absorb the excess KV cache \cite{sun2025breaking,sheng2023flexgen,chen2025impress,zheng2026solidattention}. While this solves the capacity issue, it introduces a crippling I/O bottleneck during the autoregressive decoding phase. Because dynamic sparsity methods must evaluate token importance at every decoding step, they require continuous, fine-grained retrieval of KV data from the SSD to the host. As quantitatively shown in Figure \ref{fig:2-1}, the data transfer latency from the SSD scales poorly with the context length. For instance, transferring merely 20\% to 30\% of the KV cache for token evaluation can exceed 2000 ms at a 12K context length. The systemic impact of this heavy I/O overhead is explicitly depicted in the timeline of Figure \ref{fig:2-2}. The latency required to transfer KV data across the PCIe bus completely eclipses the actual attention computation time. Consequently, the GPU experiences severe data starvation, resulting in massive idle periods that drastically reduce the end-to-end inference throughput. 

\noindent\textbf{The Inadequacy of Pure KV Compression.} Pure software-based methods attempt to fit the KV cache entirely within VRAM via aggressive quantization or heavy compression \cite{liu2024kivi,li2024snapkv,ge2024model,liu2024minicache,caipyramidkv,liu2024cachegen,hooper2024kvquant,he2024zipcache}. However, these approaches face fundamental limitations on memory-constrained PCs. First, compression only yields a constant-factor reduction (typically 25\%--60\%); the KV cache footprint still grows linearly. For example (Figure \ref{fig:context_length}), inferencing Llama-2-13B at a 32K context natively requires $\sim$240 GB. Even with aggressive INT2 compression (e.g., KIVI \cite{liu2024kivi}), the footprint remains around 96 GB. This far exceeds the capacity of commodity PCs (e.g., 24 GB VRAM + 64 GB DRAM), inevitably causing OOM failures. While dropping excess KV data avoids OOM, the resulting recomputation incurs prohibitive latency, negating the benefits of caching \cite{lee2024infinigen,zhao2024alisa}. Thus, leveraging external storage is strictly necessary for long-context inference on such devices \cite{sheng2023flexgen,sun2025breaking,chen2025impress,zheng2026solidattention}. Second, these algorithms unavoidably degrade model accuracy \cite{liu2024kivi,li2024snapkv,liu2024minicache,hooper2024kvquant}, an unacceptable trade-off for reasoning tasks. Finally, runtime compression and decompression cycles introduce further computational latency overhead.

\noindent\textbf{The Pitfalls of Existing In-Storage Inference.} 
Recent in-storage frameworks fail to support online long-context generation on standard, memory-constrained PCs. Works like InstInfer \cite{pan2025instattention} and HILOS \cite{jang2025inf} attempt to offload the entire attention computation (i.e., near-data attention) to the storage layer. However, calculating exact attention weights involves complex operations that quickly exhaust the strict compute budget of onboard FPGAs. To circumvent this, they must either rely on heavily specialized, custom-built hardware (InstInfer) or scale out to multiple SmartSSDs tailored for offline batched inference (HILOS). Consequently, these approaches inherently mismatch with the dynamic KV eviction paradigm and cannot deliver low-latency serving on a single AIPC.
\begin{figure}[tbp]
    \centering
    \includegraphics[width=0.49\textwidth]{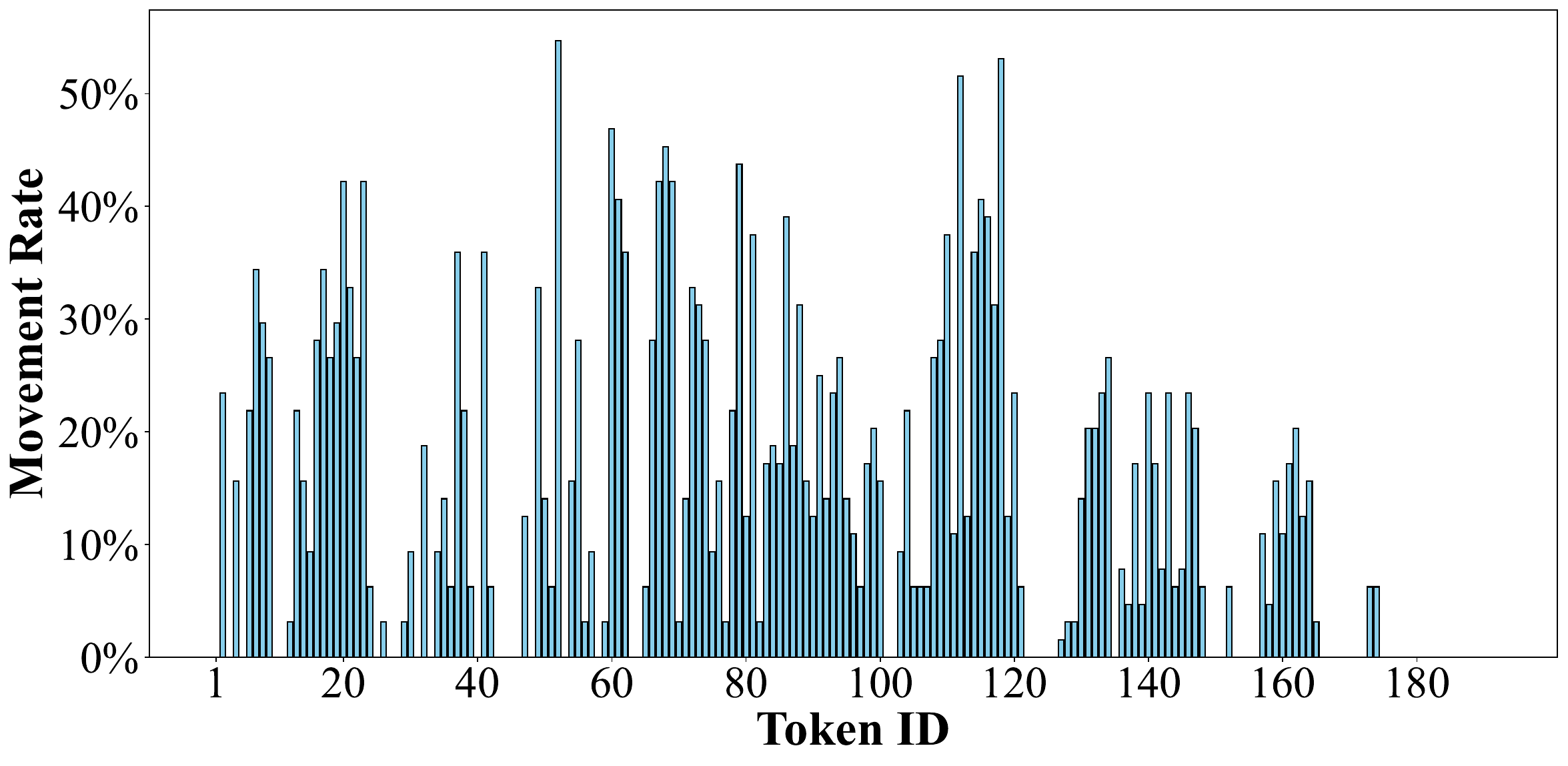}
    \caption{Percentage of Decoding Steps with KV Cache Movement using OPT-6.7b model and the Prompt in PG-19 Dataset.}
    \label{fig:token_migration}
\end{figure}

In summary, the pursuit of long-context LLM serving on memory-constrained PCs faces a series of critical roadblocks: host-centric offloading inevitably triggers OOM, pure software compression sacrifices model accuracy while merely delaying the physical memory wall, traditional SSD-assisted serving suffers from severe PCIe I/O starvation, and existing in-storage attention schemes succumb to FPGA resource exhaustion, demanding custom or multi-device hardware. To break this impasse, we propose HillInfer, an architecture that leverages a single commodity CSD (e.g., SmartSSD) to perform lightweight in-storage importance evaluation to ensure the low-latency long-context LLM inference on AIPCs.

\subsection{Challenges of HillInfer}

While leveraging Computational Storage Devices (CSDs) provides a hardware foundation to bypass the PCIe bandwidth wall, integrating them into importance-aware KV eviction introduces two profound challenges:

\noindent\textbf{Challenge 1: I/O Thrashing in Query-Dependent Eviction.} 
Token importance is inherently query-dependent \cite{sun2025breaking,lee2024infinigen,tang2024quest}, meaning it fluctuates dynamically across decoding steps. A KV data evicted to the SmartSSD in step $t$ might suddenly become critical in step $t+1$. Under such conditions, a straightforward storage management strategy will trigger severe I/O thrashing—frequent ``ping-pong'' data movement between the host DRAM and the SmartSSD. To demonstrate this visually, we distribute the KV cache across the memory hierarchy: 20\% (highly critical) pinned in GPU VRAM, 60\% in host DRAM, and 20\% in external storage. We profiled the OPT-6.7B model using the PG-19 dataset, generating 64-token outputs from 128-token prompt inputs. As shown in Figure \ref{fig:token_migration}, our profiling reveals that an average of 20\% to 50\% of the KV cache is frequently migrated back and forth between the host DRAM and external storage. This severe ping-pong I/O thrashing introduces prohibitive data transmission latency, entirely offsetting the latency benefits gained from in-storage computation.
\begin{figure}[tbp]
    \centering
    \includegraphics[width=0.49\textwidth]{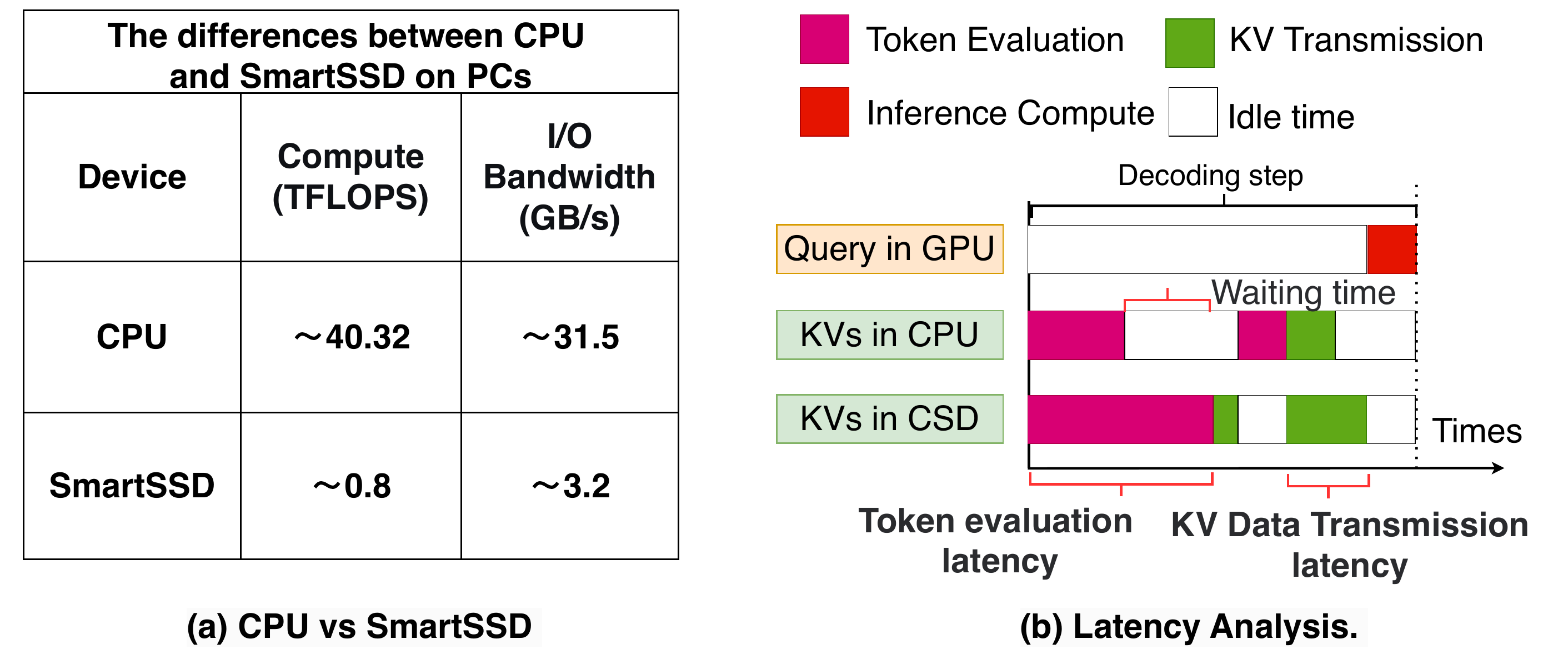}
    \caption{Heterogeneous Devices Analysis. (a) CPU vs SmartSSD: measured compute capability and I/O bandwidth. (b) The computation and transmission latency.}
    \label{fig:difference}
\end{figure}

\noindent\textbf{Challenge 2: Orchestrating a Heterogeneous Inference Pipeline.} 
While the CSD successfully eliminates massive SSD-to-host data movement during token evaluation via near-data processing, it transforms the AIPC into a highly heterogeneous environment, complicating system coordination across two critical dimensions. First, on the computation front, a severe compute capacity gap exists between the multi-core CPU and the SmartSSD's on-board FPGA. We tested the compute and I/O bandwidth capability on the AIPC, as shown in Figure \ref{fig:difference} (a). A naive division of the evaluation workload will inevitably cause synchronization stalls, as the faster processor must idle waiting for the slower one. Figure \ref{fig:difference} (b) shows the waiting time. Second, on the I/O front, a strict bandwidth disparity emerges during the final KV prefetching phase. When transmitting the selected important KV data to the GPU for attention computation, the rapid data transfers from the host DRAM clash with the slow retrievals from the SmartSSD. As depicted in Figure \ref{fig:difference} (b), failing to optimally orchestrate these heterogeneous compute and I/O operations allows the SmartSSD to dominate the entire KV transmission phase, ultimately starving the GPU and heavily penalizing end-to-end throughput.

\section{System Design of HillInfer}
In this section, we detail the designs of HillInfer. We first describe the system framework and workflow (Sec. \ref{Sec:SysOver}). Then, we introduce the hierarchical KV cache manager (Sec. \ref{Sec:HKVM}). Finally, we present an adaptive prefetch-based pipeline coordinating CPU, GPU, and SmartSSD (Sec. \ref{Sec:APP}), followed by CSD-based evaluation configuration (Sec. \ref{Sec:CEC}).

\subsection{System Overview}\label{Sec:SysOver}

We present HillInfer, an efficient in-storage KV cache eviction framework designed for long-context LLM serving on commodity PCs. Rather than treating the underlying SSD merely as a passive swap space, HillInfer shifts the paradigm towards active near-data processing. By deeply integrating commercial SmartSSDs into the inference memory hierarchy, it fundamentally decouples the query-dependent token evaluation from massive PCIe data movement, paving the way for scalable, I/O-free long-context inference.

\begin{figure}[tbp]
    \centering
    \includegraphics[width=0.49\textwidth]{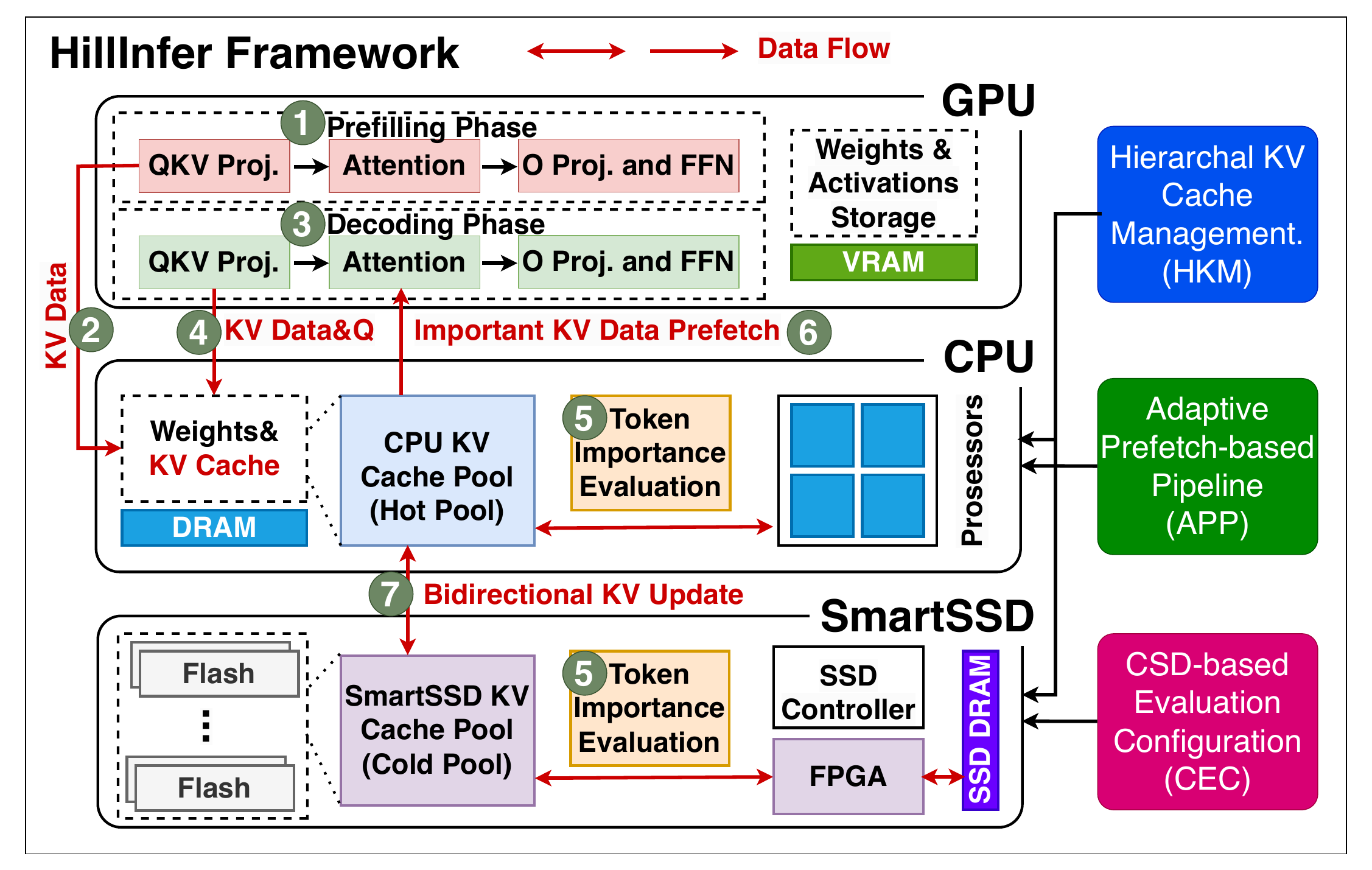}
    \caption{The Framework Overview of HillInfer.}
    \label{fig:Framework}
\end{figure}

\noindent\textbf{System Framework.} As illustrated in Figure \ref{fig:Framework}, HillInfer transforms a commodity PC into a tightly coupled, three-tier heterogeneous computing architecture. At the top tier, the GPU acts as the primary execution engine, housing the LLM weights and activation. It is strictly reserved for compute-intensive attention operations, caching only the most critical or newly generated KV data. The host CPU and its DRAM serve as the central orchestrator and the hot cache tier. The host DRAM maintains the CPU KV Cache Pool to buffer tokens exhibiting high access locality, while the processors manage local token evaluation, global score merging, and pipeline scheduling. At the bottom tier, the SmartSSD absorbs the massive long-tail KV data that exceeds host memory limits. By utilizing its on-board NAND as a cold cache pool and its embedded FPGA for near-data processing, the SmartSSD evaluates token importance directly at the storage level, fundamentally bypassing the host PCIe bus.

\noindent\textbf{System Workflow.} The inference data flow of HillInfer is meticulously orchestrated to maximize parallel execution and minimize redundant data movement. During the prefilling phase (Steps 1 and 2 in Figure \ref{fig:Framework}), the GPU processes the dense prompt, and the generated massive KV data are systematically offloaded and distributed across the CPU and SmartSSD cache pools to prevent VRAM exhaustion. In each subsequent autoregressive decoding step (Step 3), the GPU generates a new Query ($Q$) vector along with the KV data of the newly generated token. Instead of pulling massive historical KV data up to the GPU, HillInfer broadcasts the lightweight $Q$ vector down to both the CPU and the SmartSSD (Step 4). The CPU and the SmartSSD FPGA then independently and concurrently evaluate the importance of their locally stored KV data against the broadcasted $Q$ (Step 5). Crucially, the SmartSSD scores its data internally, ensuring no raw KV data traverse the PCIe bus during this intensive evaluation phase. Subsequently, the SmartSSD returns only a lightweight vector of importance scores to the host. The CPU aggregates these scores globally, identifies the most critical tokens, and orchestrates the prefetching of only these crucial KV data back to the GPU VRAM for the actual attention computation (Step 6). Finally, to prevent I/O thrashing caused by dynamic query dependencies, HillInfer continuously executes a bidirectional KV update mechanism (Step 7), dynamically promoting highly activated tokens from the SmartSSD to the host DRAM and demoting cold tokens to the storage, ensuring the most frequently accessed data remains close to the CPU.

In this system workflow, to quantify token importance, HillInfer adopts a widely validated metric building upon established sparsity paradigms \cite{zhang2023h2o,lee2024infinigen,sun2025breaking}, where a historical token's importance is defined by the column-wise summation of its corresponding attention weights.

\noindent\textbf{HillInfer Components.} As shown in Figure \ref{fig:Framework}, there are three major components in the HillInfer runtime. The first is the Hierarchal KV Cache Management (HKM). It continuously monitors and manages the bidirectional KV cache pools distributed across the CPU memory and the SmartSSD to maintain data locality and prevent I/O thrashing, which we discuss in Section \ref{Sec:HKVM}. The second component is the Adaptive Prefetch-based Pipeline (APP). This module orchestrates the asynchronous execution and heterogeneous data transfers among the GPU, CPU, and SmartSSD to achieve optimal I/O-compute overlap. We explain its scheduling operations in Section \ref{Sec:APP}. Additionally, to perform near-data processing efficiently, the CSD-based Evaluation Configuration (CEC) optimizes the token importance evaluation workload to fit the strict hardware constraints of the SmartSSD FPGA. We detail this hardware-level configuration in Section \ref{Sec:CEC}.

\subsection{Hierarchal KV Cache Management.}\label{Sec:HKVM}
\begin{figure}[tbp]
    \centering
    \includegraphics[width=0.49\textwidth]{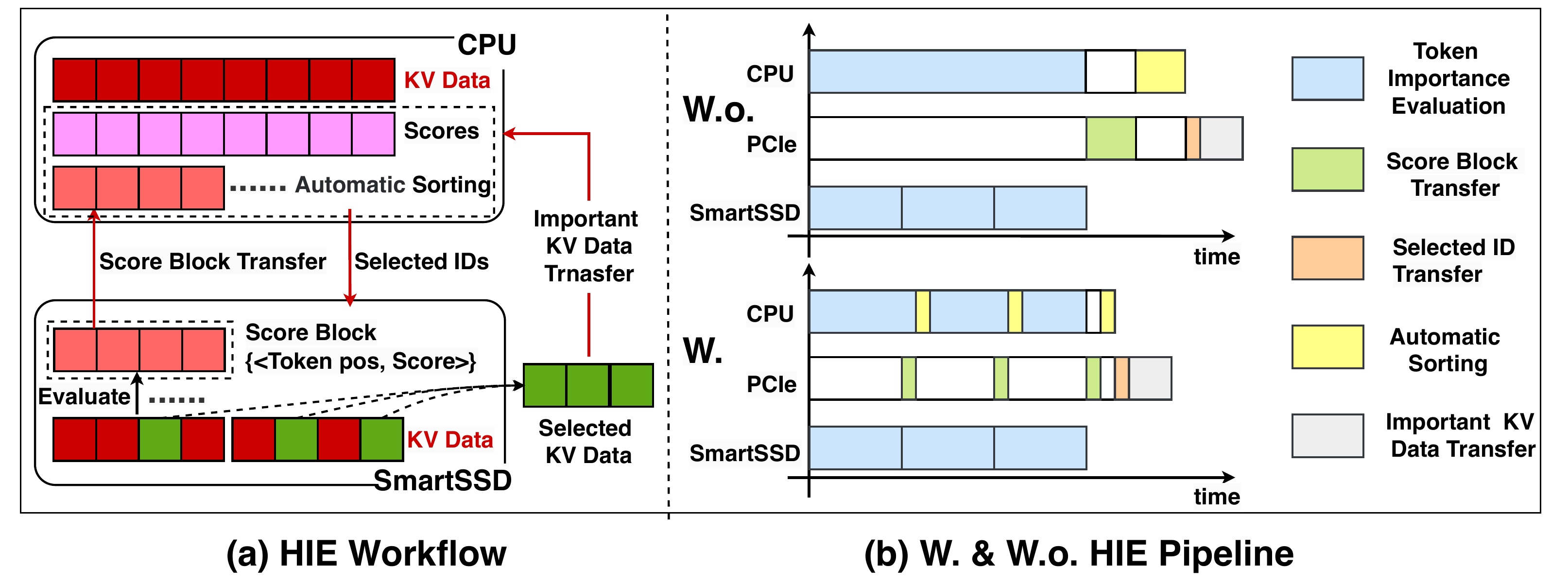}
    \caption{The Workload and Pipeline of HIE.}
    \label{fig:HIE}
\end{figure}

\noindent\textbf{Hierarchical Importance Evaluation (HIE).} 
While offloading evaluation to the SmartSSD fundamentally mitigates massive KV data movement, it introduces a distributed synchronization bottleneck. If the host CPU waits for the SmartSSD to finish evaluating its entire local KV pool before retrieving the scores, the system suffers from severe synchronization stalls and unhidden PCIe latency, as depicted in the serialized ``W.o.'' timeline in Figure \ref{fig:HIE}(b). To eliminate these pipeline bubbles, HillInfer implements a fine-grained, asynchronous workflow. As illustrated in Figure \ref{fig:HIE}(a), rather than adopting a monolithic execution model, the SmartSSD processes its local KV pool in chunks. Upon evaluating a batch of $n$ tokens, the FPGA immediately packs the results into a compact \textit{Score Block} (formatted as lightweight $<$\textit{Token pos, Score}$>$ tuples) and streams it to the host via the PCIe bus. Crucially, the FPGA proceeds to evaluate the next batch without halting. 

Concurrently, the CPU continuously receives these incoming Score Blocks and asynchronously merges them with its own locally evaluated scores. As shown in the ``W.'' timeline of Figure \ref{fig:HIE}(b), this chunk-based streaming mechanism perfectly overlaps the Score Block transfer (green blocks) and the CPU's automatic sorting (yellow blocks) with the ongoing token evaluation (blue blocks). By pipelining these heterogeneous operations, HIE completely hides the cross-device communication overhead. Once the global top-$\alpha$ percent indices are resolved, the CPU selectively fetches only the critical KV data from the SmartSSD, achieving near-perfect I/O-compute overlap and minimizing the end-to-end evaluation latency.

\noindent\textbf{Insights: Temporal Locality and Stable Hit Rates.} Because token importance is inherently query-dependent at each decoding step \cite{sun2025breaking,lee2024infinigen,tang2024quest}, naive eviction strategies that strictly adhere to these fluctuating scores often trigger excessive ping-pong data movement (I/O thrashing) between the host memory and the storage tier. To demystify these dynamic access patterns, we comprehensively profile token eviction dynamics across diverse LLMs and datasets. Specifically, we track the lifetime access frequencies of individual tokens throughout the entire autoregressive decoding phase, recording the exact number of times each token is dynamically classified as \textit{Important} (top 20\%), \textit{Secondary} (middle 60\%), or \textit{Unimportant} (bottom 20\%). As a representative example illustrated in Figure \ref{fig:Importance_count}, our profiling reveals two critical access patterns that can fundamentally mitigate this bottleneck.

First, token importance demonstrates pronounced \textit{temporal locality}. As observed in the recently generated tokens (the rightmost distribution in Figure \ref{fig:Importance_count}), newly produced KV data have an extremely high probability of remaining critical for the immediately following decoding steps.

Second, token access exhibits strong \textit{hit-rate stability}. A distinct subset of historical tokens consistently maintains high importance across almost all decoding steps, remaining heavily concentrated in the Important tier (depicted by the tall red bars). These tokens act as persistent ``attention sinks'' and inherently possess a remarkably high cache hit rate.

Driven by these dual insights, we propose the Bidirectional KV Cache Pools (BKP). By strategically pinning these historically high-hit-rate and temporally recent tokens in the fast host DRAM, BKP absorbs the vast majority of dynamic accesses locally, thereby fundamentally suppressing I/O thrashing.

\noindent\textbf{Bidirectional KV Cache Pools (BKP).} 
Built upon the aforementioned insights, HillInfer introduces Bidirectional KV Cache Pools (BKP). As depicted in Figure \ref{fig:Framework}, the memory hierarchy is bifurcated into a \textit{Hot Pool} (CPU KV Cache Poll) and a \textit{Cold Pool} (SmartSSD KV Cache Pool). Our primary objective is to maximize the local hit rate within the Hot Pool, thereby fundamentally eradicating I/O thrashing, that is, the frequent ping-pong data movement of historically critical tokens across the heterogeneous CPU-SmartSSD boundary. To achieve this, BKP implements a dual-metric placement and offloading strategy:
\begin{figure}[tbp]
    \centering
    \includegraphics[width=0.49\textwidth]{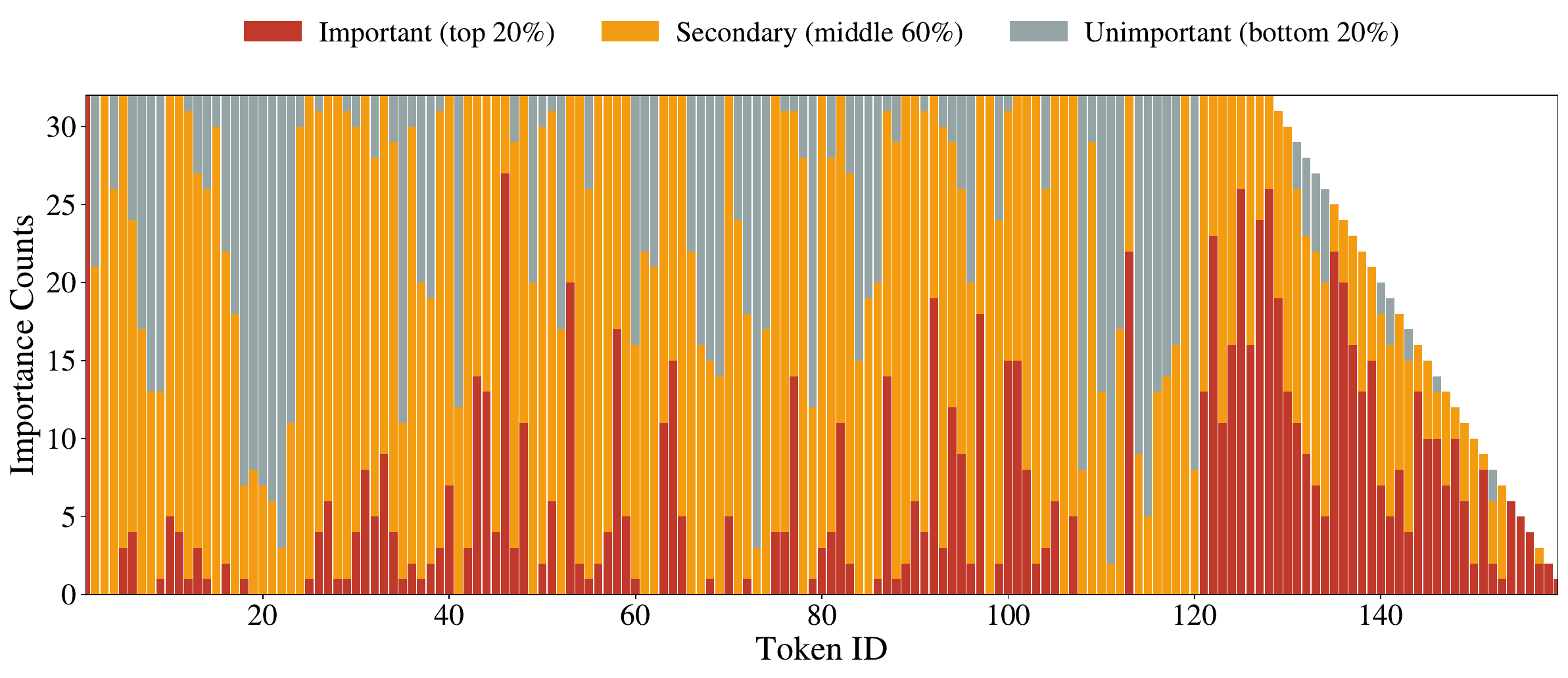}
    \caption{The Importance Count in the Whole Decoding Stage: 128 prompt length, 32 outputs, opt-6.7b model, and PG-19 datasets.}
    \label{fig:Importance_count}
\end{figure}

\textit{(1) Temporal-Aware Pinning:} To exploit temporal locality \cite{zhao2024alisa}, BKP enforces a strict recent-token pinning policy. Recognizing that newly generated tokens have an overwhelmingly high probability of being attended to in immediately subsequent steps, we statically lock the KV cache of the most recent $\alpha \cdot N$ steps exclusively in the Hot pool, where $N$ denotes the current context length and $\alpha$ is a user-configurable capacity ratio (typically set between $10\%$ and $20\%$ in our evaluation). By proactively isolating these volatile new tokens within the upper (host) memory hierarchy, this temporal pinning strategy strictly shields the PCIe bus and the storage tier from massive transient I/O requests.

\textit{(2) Frequency-Aware Tracking:} Beyond recency, some older tokens act as persistent ``attention sinks'' and exhibit remarkably stable cache hit rates. To capture these, BKP maintains a lightweight, globally synchronized KV Cache Hit-Rate Table. By dynamically tracking access frequencies across decoding iterations, BKP guarantees that the top-$\alpha$ fraction of globally critical KV tensors is permanently retained in the Hot Pool of host DRAM, shielding them from being offloaded to the SmartSSD.

\textit{(3) Bidirectional Promotion and Demotion:} Finally, to adapt to phase shifts in long-context inference, BKP continuously executes a bidirectional state update as a background process after each decoding step. During the maintenance of the Hit-Rate Table, if a cold token residing on the SmartSSD experiences a sudden surge in importance and breaches the top-$\alpha$ threshold, BKP triggers a cache promotion, immediately migrating it up to the CPU Hot Pool. Conversely, to strictly bound the host memory footprint, an equal volume of the lowest-ranked, decaying KV tensors in the CPU Hot pool suffers cache demotion and is asynchronously evicted down to the SmartSSD. This bidirectional swapping ensures the Hot Pool is consistently populated with the highest-utility data, systematically silencing the detrimental ping-pong I/O thrashing across the PCIe bus.

\subsection{Adaptive Prefetch-based Pipeline (APP)}\label{Sec:APP}

To perfectly mask the overhead of KV cache management, HillInfer builds upon layer-wise prefetching paradigms \cite{lee2024infinigen,sun2025breaking,chen2025impress}, strategically overlapping the token importance evaluation and KV data prefetching of layer $i{+}1$ with the GPU's inference computation of layer $i$. However, realizing a seamless pipeline in a memory-constrained, heterogeneous environment is non-trivial. 

\begin{figure}[tbp]
    \centering
    \includegraphics[width=0.48\textwidth]{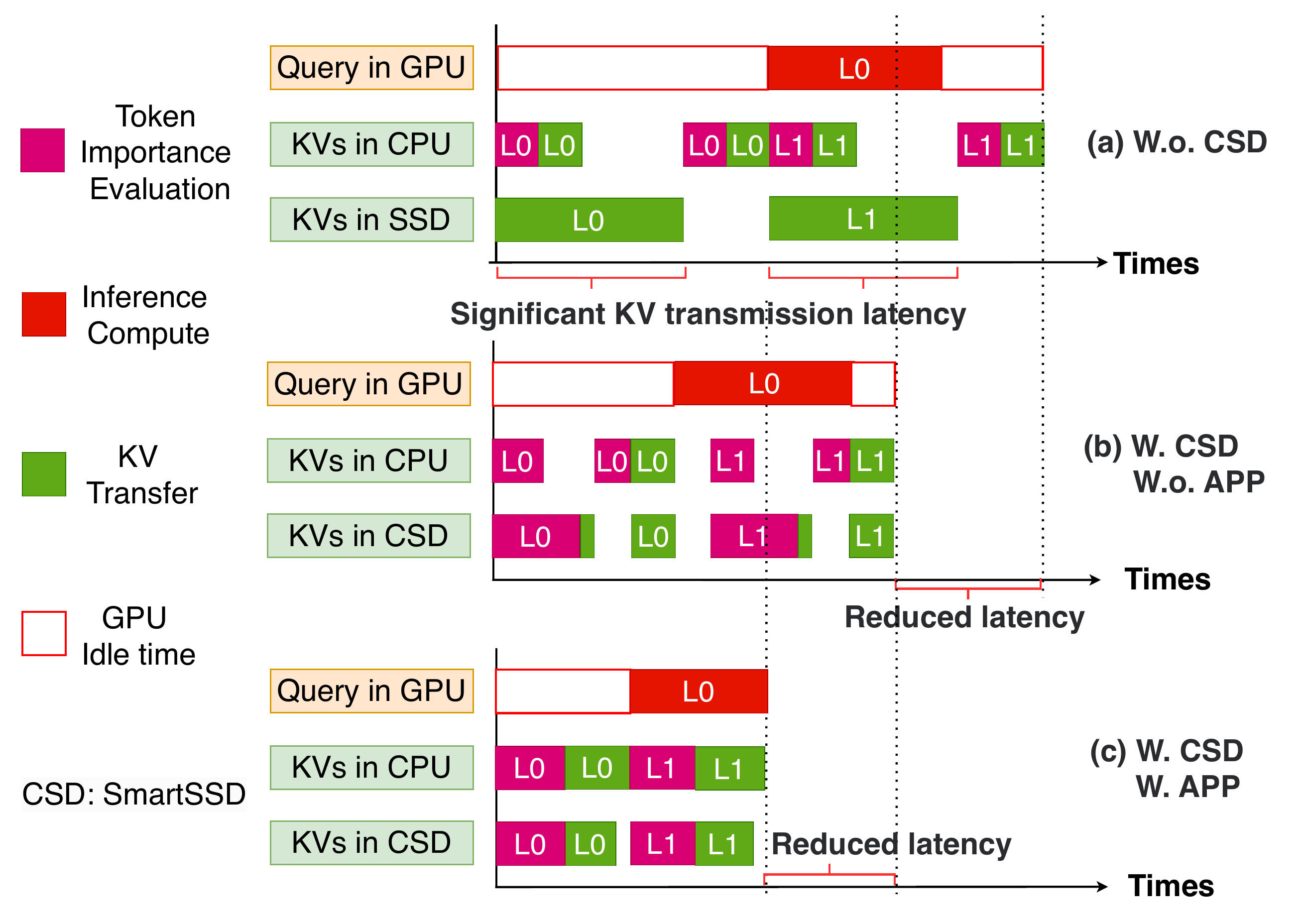}
    \caption{The Latency Comparison without CSD and with CSD and APP Technique.}
    \label{fig:APP}
\end{figure}

As illustrated in Figure \ref{fig:APP}(a), under traditional SSD-based eviction schemes, the massive raw KV data transfers from the external storage to the host dictate the critical path, inducing significant transmission latency and severe GPU starvation. While integrating a SmartSSD effectively eliminates this raw I/O bottleneck via near-data processing (Figure \ref{fig:APP}(b)), it simultaneously introduces a strict synchronization barrier. Because the multi-core CPU and the embedded FPGA possess drastically different processing throughputs (Figure \ref{fig:difference}), naively partitioning the KV cache pools between them inevitably causes a straggler effect. The faster processor finishes its evaluation early and stalls, while the slower processor delays the global score aggregation. Consequently, this uncoordinated heterogeneous execution fails to hide the latency, leaving the GPU idle and heavily degrading the end-to-end throughput. 

To overcome this, we propose the Adaptive Prefetch-based Pipeline (APP) to dynamically balance the distributed workload and achieve optimal I/O-compute overlap (Figure \ref{fig:APP}(c)). To fundamentally eliminate the straggler effect, APP formulates an analytical latency model to adaptively determine the optimal capacity ratio $\beta$ between the two hierarchical KV cache pools.

\noindent\textbf{Analytical Latency Model.} 
Assume the total volume of KV tensors for the current query is $M$, partitioned into $M_c$ (stored in the CPU Hot Pool) and $M_s$ (stored in the SmartSSD Cold Pool). Let $f_c$ and $f_s$ denote the average data processing throughput of the CPU and the SmartSSD FPGA, respectively. Let $B_c$ (e.g., PCIe Gen4$\times$16) and $B_s$ (e.g., PCIe Gen4$\times$4) represent their respective effective PCIe transmission bandwidths. Finally, let $\alpha$ be the target token retention ratio (i.e., the proportion of selected important KV data) and $M_0$ be the strict upper bound of the available host memory allocated for the KV cache. The end-to-end latency of the evaluation and prefetching phase is dictated by the maximum processing time of the two devices. To achieve perfect synchronization and minimize GPU idle time, the latency of the CPU path ($T_{CPU}$) must perfectly align with that of the SmartSSD path ($T_{SmartSSD}$):
\begin{equation}
T_{CPU} = \frac{M_c}{f_c} + \frac{\alpha M_c}{B_c} \approx \frac{M_s}{f_s} + \frac{\alpha M_s}{B_s} = T_{SmartSSD}
\end{equation}
Subject to the strict host memory constraint:
\begin{equation}
M_c \leq M_0.
\end{equation} 
By solving this balance equation, APP dynamically orchestrates the cache capacity ratio $\beta$ as follows:
\begin{equation}
\beta = \frac{M_c}{M_s} \approx \frac{B_c f_c (B_s + \alpha f_s)}{B_s f_s (B_c + \alpha f_c)}, \quad \left( s.t. \ \beta \leq \frac{M_0}{M_s} \right)
\end{equation}

where the hardware-specific constants $f_c, f_s, B_c,$ and $B_s$ are acquired through offline profiling at initialization. This derived condition guarantees that both devices complete their evaluation and data transmission in a nearly synchronized manner, seamlessly hiding the KV prefetching latency behind the GPU's attention computation (as shown in Figure \ref{fig:APP}(c)).

\noindent\textbf{Empirical Skewness Correction.} 
In practice, since HKM deliberately pins high-hit-rate tokens in the host DRAM, the actual ratio of important KV data in the CPU slightly exceeds the theoretical estimation ($>\alpha M_c$), while the SmartSSD holds a colder distribution ($<\alpha M_s$). To prevent this inherent data skewness from breaking the pipeline synchronization, APP applies a lightweight empirical correction to $\alpha$ when calculating the optimal $\beta$, ensuring robust load balancing.

\subsection{CSD-based Evaluation Configuration}\label{Sec:CEC}

As shown in Figure \ref{fig:difference}(a), the SmartSSD's onboard FPGA (e.g., Xilinx KU15P) operates under strict resource constraints compared to the host CPU. Directly porting host-centric evaluation logic to the CSD is highly inefficient, necessitating a hardware-aware computational configuration.

\noindent\textbf{Algorithmic Simplification.} 
Existing host-centric frameworks typically evaluate token importance using exact attention weights, computing $\rho(Q \cdot K^T / \sqrt{d})$ \cite{zhang2023h2o,zhao2024alisa}. However, since KV eviction only requires \textit{ranking} historical tokens, and the Softmax function $\rho(\cdot)$ is strictly monotonic, HillInfer strips all complex exponentiation and scaling division operations. The FPGA is configured to compute only the raw inner product ($Q \cdot K^T$). This mathematical reduction completely eliminates the need for resource-heavy transcendental function units and hardware dividers, drastically minimizing the logic utilization footprint and freeing up critical on-chip resources for the core dot-product operations \cite{samsungSmartSSD}.

\noindent\textbf{Streaming Execution and Asymmetric Precision.} 
To execute these inner products at maximum throughput, HillInfer pins the lightweight $1 \times d$ Query vector into the FPGA's BRAM and continuously streams the $N \times d$ Key matrix. By utilizing a fully unrolled Adder Tree pipeline, the FPGA effectively hides the memory access latency behind the parallelized computation \cite{wang2021spatten,samsungSmartSSD}. Crucially, to further alleviate the severe bandwidth constraints and logic utilization, we introduce a \textit{scoring-specific asymmetric precision} mechanism. Unlike pure-software quantization that permanently degrades tensor values \cite{liu2024kivi,hooper2024kvquant}, HillInfer performs the $Q \cdot K^T$ operations using low-precision arithmetic (e.g., INT8/INT4) strictly for the evaluation phase, dynamically casting the streamed FP16 Keys on the fly. Once the top-$K$ indices are identified, the SmartSSD retrieves and transmits the original, uncompressed FP16 KV tensors back to the GPU. This decoupled design halves resource consumption without incurring any degradation in the LLM's final accuracy.

\section{Implementation and Experiments}
\subsection{Implementation}
We implement HillInfer by extending the offloading-based LLM inference framework Flex \cite{sheng2023flexgen} with 500+ lines of additional Python code, while incorporating some codes about data movement from the prefetch-based framework InfiniGen \cite{lee2024infinigen}. Furthermore, we developed HLS-based C++ code to implement token-level KV cache importance evaluation on the FPGA units of the SmartSSD. Host–SmartSSD communication is implemented using the Xilinx Runtime (XRT) interface \cite{xilinxXRT}. Our preliminary experiments are built upon the Hugging Face Transformers framework \cite{wolf2019huggingface}, and HillInfer can be readily applied to accelerate most open-source LLMs, such as LLaMA \cite{huang2023lawyer}, Qwen \cite{bai2023qwen}, and OPT, etc., without requiring modifications to model architectures.

\subsection{Experiments Setting}
\begin{figure}[htbp]
    \centering
    \includegraphics[width=0.38\textwidth]{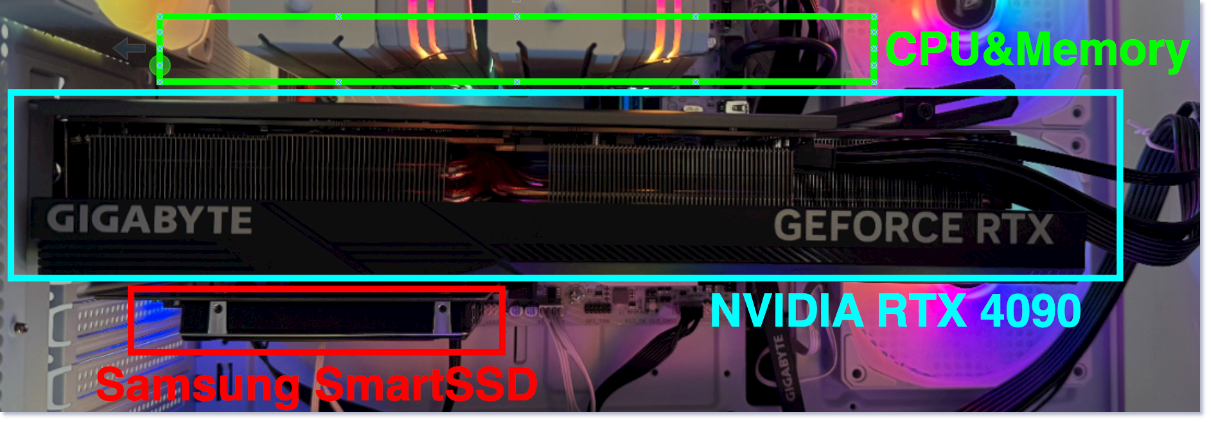}
    \caption{Hardware Architecture of HillInfer.}
    \label{fig:hardware}
\end{figure}

\noindent\textbf{Hardware.}
To closely reflect realistic PC deployment scenarios, we conduct our experiments on a Ubuntu 22.04 desktop system equipped with an NVIDIA GeForce RTX 4090 GPU, an Intel(R) Xeon(R) Platinum 8352V CPU @ 2.10GHz, 24 GB of GPU memory, 64 GB of host memory, and a 300 GB Intel SSD connected via a PCIe 4.0 interface. We further employ a Samsung SmartSSD as the computational storage device with Xilinx’s UltraScale+ FPGA, 4 GB DDR4, and 4 TB NAND Flash, providing a peak bandwidth of approximately 3.2 GB/s, which is connected to the motherboard through PCIe (3.0 interface), as illustrated in Figure \ref{fig:hardware}.
\begin{figure*}[tbp]
    \centering
    \includegraphics[width=\textwidth]{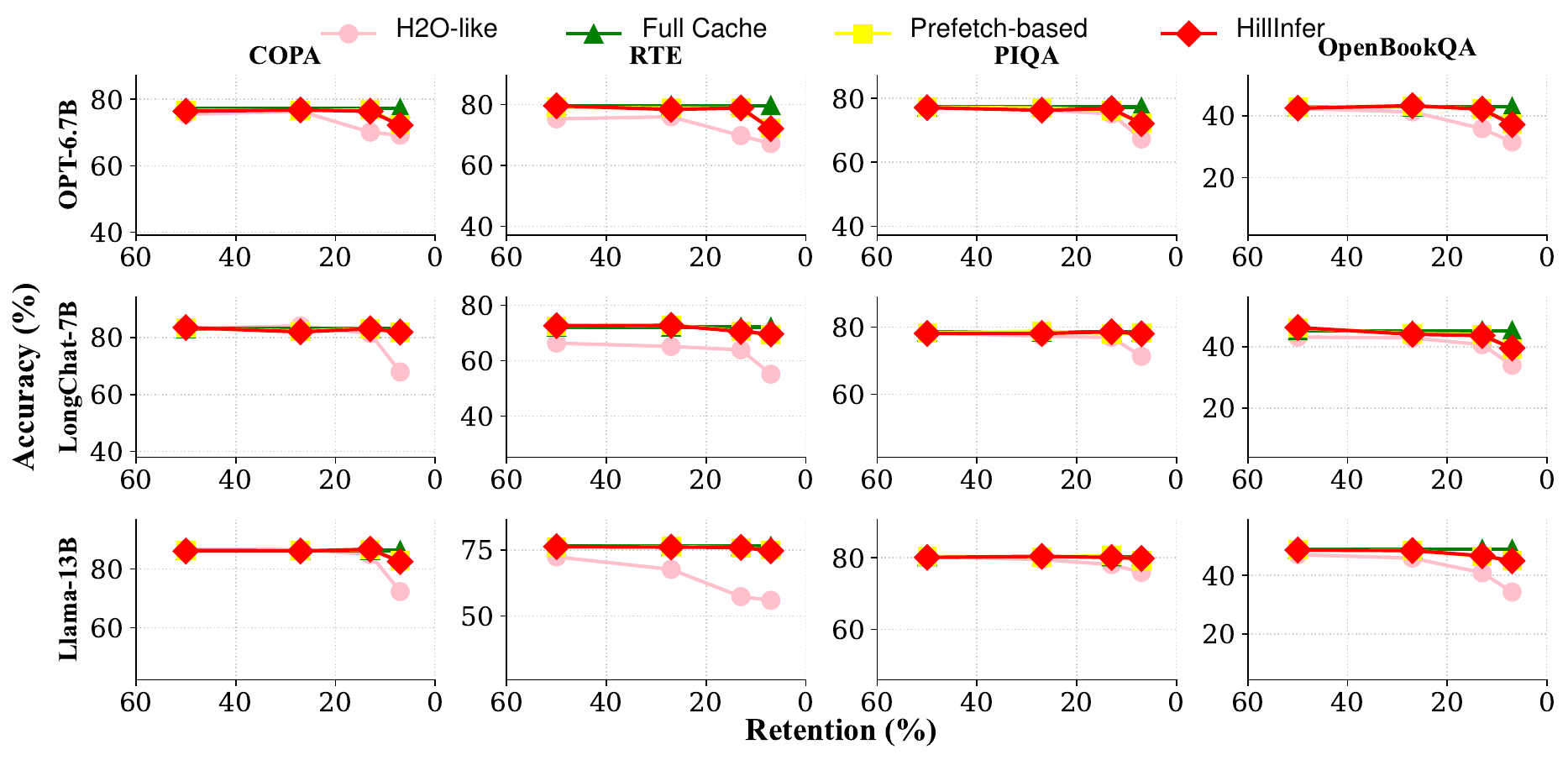}
    \caption{The accuracy comparison of different frameworks across four datasets and three models under different relative KV Cache sizes.}
    \label{fig:accuracy}
\end{figure*}

\noindent\textbf{Models.} 
To evaluate the adaptability of HillInfer across diverse model architectures and scales, we conduct experiments using models from the LLaMA family \cite{huang2023lawyer} (LongChat-7B and LLaMA-13B), Qwen-7B \cite{bai2023qwen}, and OPT-6.7B. 


\noindent\textbf{Datasets.} For evaluating model inference accuracy, we use several few-shot tasks from the LM-evaluation-harness benchmark \cite{gao2021framework}, including OpenBookQA, RTE, PIQA, COPA, and PG-19 \cite{rae2019compressive}. To assess long-context inference latency, we conduct experiments using LongBench \cite{bai2024longbench} with the context length up to 36K.

\noindent\textbf{Baselines.} We compare HillInfer against four baseline approaches.	(1) Full Cache: This baseline retains the entire KV cache without performing any token importance evaluation and is commonly treated as the accuracy-test baseline. (2) H2O-like: This approach adopts a token importance evaluation strategy similar to H2O \cite{zhang2023h2o} and employs an offloading-based mechanism for KV cache transfer. (3) Prefetch-based \cite{lee2024infinigen}: This baseline follows an InfiniGen-style design, performing importance evaluation and prefetching layer-wise KV data from the external storage and host memory. (4) LeoAM-like: This approach applies the LeoAM method \cite{sun2025breaking} for token importance evaluation and leverages the KV-Abstract to optimize KV data transfer.

\noindent\textbf{Comparison metics.} For model accuracy evaluation, we use Accuracy (\%) as the metric. End-to-end inference performance is measured using Latency (ms). To assess throughput and acceleration effectiveness, we report Speedup ($\times$), respectively. All evaluation metrics follow those adopted in prior work \cite{lee2024infinigen,chen2025impress,sun2025breaking}. We refer to H2O \cite{zhang2023h2o} and set the importance rate $\alpha$ as 20\% and batch size = 8 in our experiment. KV Cache are store under standard 16-bit floating-point (FP16) precision.
\begin{figure*}[tbp]
    \centering
    \includegraphics[width=0.8\textwidth]{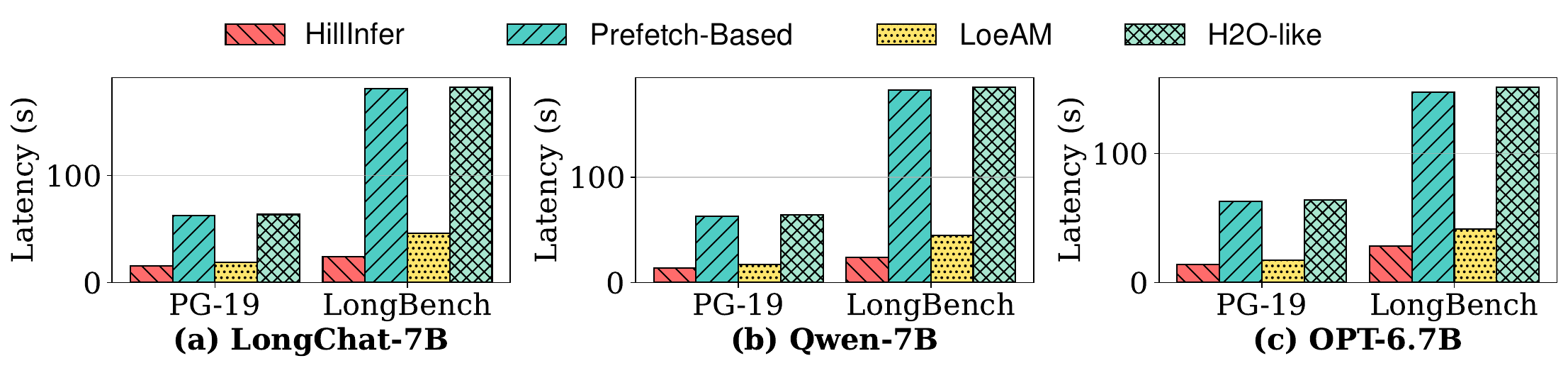}
    \caption{The Inference Latency Comparison with Different Models on PG-19 and LongBench.}
    \label{fig:latency}
\end{figure*}
\begin{figure*}[tbp]
    \centering
    \includegraphics[width=0.8\textwidth]{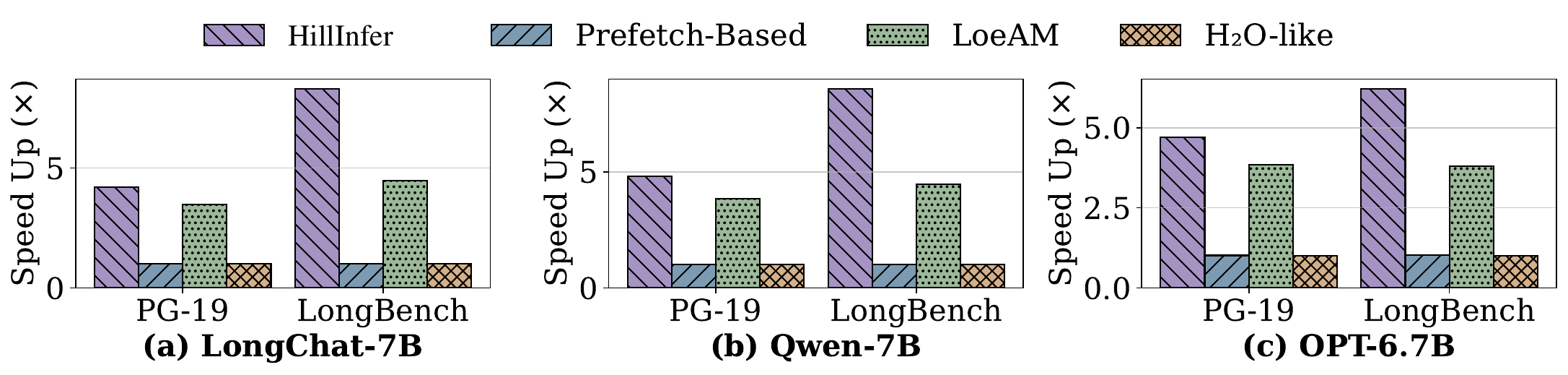}
    \caption{The Speed Up Comparison with Different Models on PG-19 and LongBench.}
    \label{fig:SpeedUp}
\end{figure*}
\subsection{Main Results}
\noindent \textbf{Model Accuracy.} Since LeoAM and prefetch-based methods achieve comparable accuracy \cite{sun2025breaking}, for simplicity, we evaluate the accuracy of HillInfer across different models and datasets against Full Cache, H2O-like, and prefetch-based methods. Our results show that HillInfer maintains comparable accuracy to these baselines, as shown in Figure \ref{fig:accuracy}. This demonstrates that, while leveraging SmartSSD for inference acceleration, HillInfer carefully designs HKM and APP to manage KV data offloading rather than incorrectly evicting KV data, thereby avoiding any degradation in model accuracy. In addition, we observe that setting the importance rate between 10\% and 20\% generally yields a favorable trade-off between performance and accuracy.

\noindent \textbf{End-to-end Latency and Throughput.} We evaluate the end-to-end latency and throughput of different models and batch sizes using LongBench. Compared to the baselines, we find that HillInfer reduces end-to-end latency by $76.25\%–88.32\%$, as shown in Figure \ref{fig:latency}. We also observe that HillInfer achieves a speedup of $4.21\sim8.56\times$, as shown in Figure \ref{fig:SpeedUp}, with more pronounced speed-up gains on datasets with longer context lengths. We scale the batch size up to 12 for OPT-6.7B to equivalently evaluate its performance under long-context workloads. Furthermore, HillInfer demonstrates universal architectural compatibility across diverse modern attention mechanisms, encompassing Multi-Head Attention (MHA), Multi-Query Attention (MQA), and Grouped-Query Attention (GQA). This inherent adaptability stems from the fact that our framework orchestrates KV cache eviction strictly at the coarse-grained token level. By decoupling the token importance evaluation from exact attention head mappings, HillInfer remains fundamentally insensitive to the variations in underlying attention topologies, ensuring broad applicability across a wide spectrum of modern LLM architectures.
\begin{figure}[tbp]
    \centering
    \includegraphics[width=0.38\textwidth]{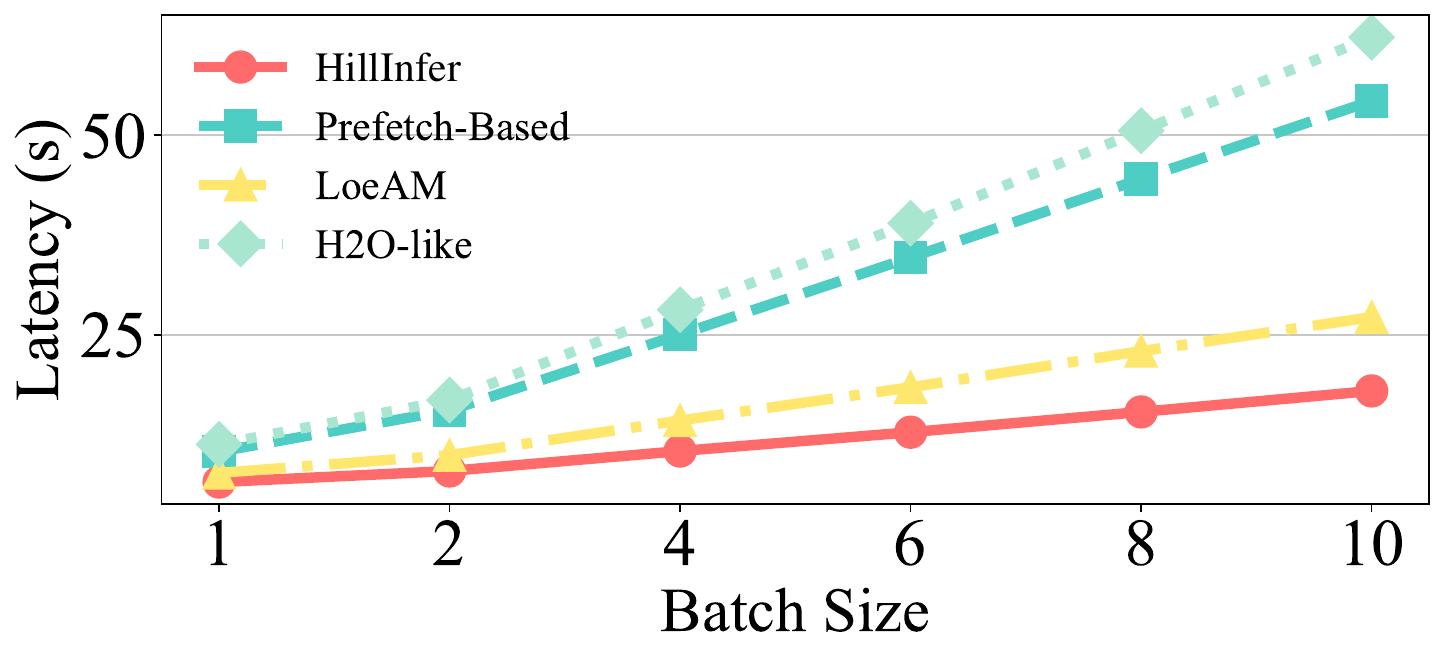}
    \caption{Latency Comparison over Different Batch Size on OPT-6.7B.}
    \label{fig:Latency_batch}
\end{figure}
\noindent\textbf{Performance Over Batch Sizes.}
We use the same setting as InfiniGen \cite{lee2024infinigen} with the sequence length of 2048 (1920 input and 128 output tokens). Figure \ref{fig:Latency_batch} illustrates inference latency as the batch size scales from 1 to 10. While latency naturally increases for all methods due to expanding KV cache footprints, traditional host-centric approaches (H2O-like and Prefetch-Based) exhibit a steep, near-linear degradation, exceeding 50s at a batch size of 10 due to severe PCIe I/O congestion. In stark contrast, HillInfer demonstrates exceptional scalability. By processing token evaluation entirely in-storage, HillInfer effectively absorbs the massive memory pressure and hides I/O overhead. Consequently, it maintains latency well under 25s even at the maximum batch size, significantly outperforming all baselines, including LoeAM.

\subsection{System Analysis}
\begin{figure}[tbp]
    \centering
    \begin{minipage}{0.23\textwidth}
        \centering
        \includegraphics[width=\textwidth]{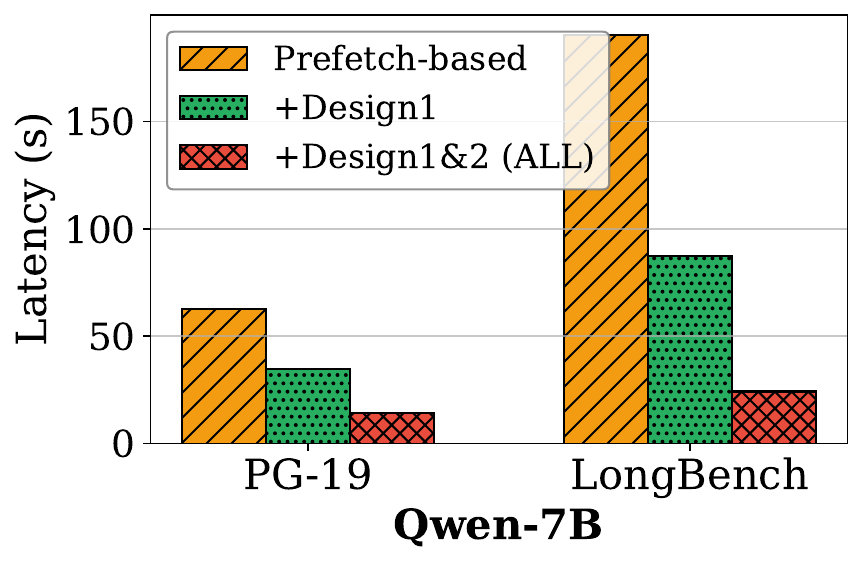}
        \subcaption{Ablation Study with LongBench \& PG-19 on Qwen-7B.}
        \label{fig:ablition}
    \end{minipage}
    \hfill
    \begin{minipage}{0.24\textwidth}
        \centering
        \includegraphics[width=\textwidth]{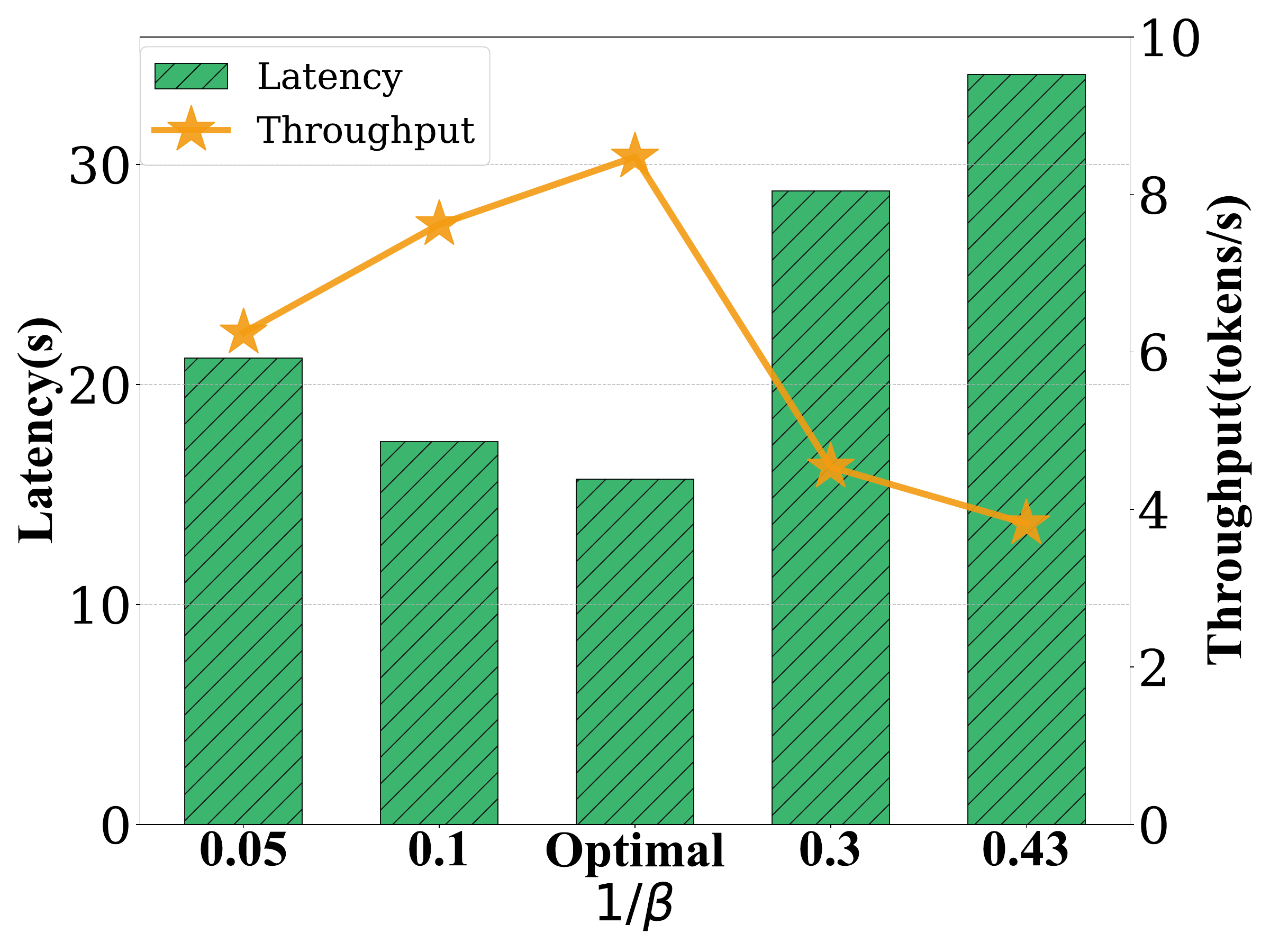}
        \subcaption{Sensitive Analysis with PG-19 on Qwen-7B.}
        \label{fig:sensitive}
    \end{minipage}
    \caption{Ablation Study and Sensitive Analysis.}
    \label{fig:Sys_Ana}
\end{figure}

\noindent\textbf{Ablation Study.} Figure \ref{fig:ablition} illustrates the individual contributions of Design 1 (HKM, Sec. \ref{Sec:HKVM}) and Design 2 (APP, Sec. \ref{Sec:APP}) to HillInfer (Note that CEC is excluded from this breakdown as it serves as a static hardware prerequisite rather than a tunable policy). Compared to the prefetch-based baseline, integrating Design 1 (+HKM) and subsequently combining both (+HKM \& APP) yield progressive reductions in inference latency. This indicates that both designs independently and synergistically contribute to accelerating the overall inference process. 

\begin{figure}[tbp]
    \centering
    \includegraphics[width=0.45\textwidth]{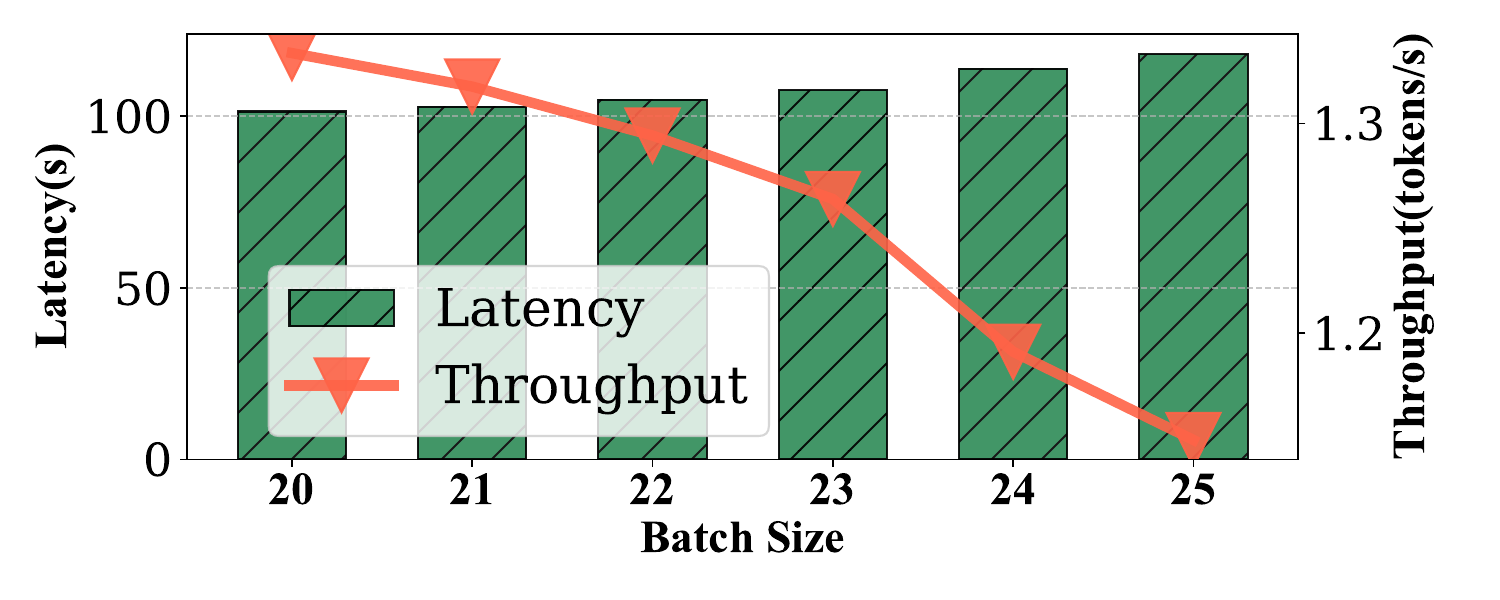}
    \caption{Latency and Throughput over Different Larger Batch Sizes on OPT-6.7B.}
    \label{fig:Latency_throughput_batch}
\end{figure}

\noindent\textbf{Sensitivity Analysis.} 
Following the evaluation protocol of InfiniGen \cite{lee2024infinigen}, we configure the workload to 1920 input tokens and 128 output tokens. Figure \ref{fig:Latency_throughput_batch} illustrates the latency and throughput of HillInfer on OPT-6.7B under extreme batch sizes (from 20 to 25). This specific setup serves as an aggressive memory pressure test, explicitly evaluating the system's robustness. Furthermore, Figure \ref{fig:sensitive} depicts the end-to-end latency variation across different tuning values of $\beta$. We observe that the latency is strictly minimized when $$
\beta \approx \frac{M_c}{M_s} \approx \frac{f_c (B_s B_c + \alpha B_c f_s + \alpha B_sf_s)}{B_s f_s (B_c + \alpha f_c)}.
$$
This empirical optimum perfectly aligns with the theoretical design rationale of our Adaptive Prefetch-based Pipeline (APP), confirming its capability to dynamically perfectly balance the heterogeneous evaluation workloads between the CPU and the SmartSSD. 

\noindent\textbf{Overhead Analysis.} In HIE, HillInfer transfers Score Blocks from the SmartSSD to the CPU. Each score block only occupies $2n$ half-precision floating-point values (i.e., $4n$ bytes), which is negligible compared to the memory footprint of the KV cache. As a result, its transfer latency can be fully hidden by computation. In BKP, HillInfer maintains a cache hit table with a size of $2N$ bytes, which is negligible compared to the memory footprint of the KV cache.

\section{Related Work}

\noindent\textbf{Long-context LLM Inference Framework.} A large number of recent works study long-context LLM serving in data center networks, with an emphasis on resource scheduling and KV cache management to improve system throughput and inference latency \cite{agrawal2024medha,qin2025mooncake,hooper2024kvquant,lim2024accelerating,lin2024infinite,wu2024loongserve}. Many existing works also focus on long-context LLM inference on single-node GPU with large memory capacity (e.g., A100), demonstrating that inference latency can be significantly reduced through algorithm–system co-design \cite{liu2024cachegen,pan2025instattention,yao2025cacheblend,zhao2024alisa,zhang2025pqcache}. However, for AIPCs with limited resources, long-context LLM inference is fundamentally constrained by memory capacity. Some prior works explore leveraging external memory, such as SSDs, to accelerate inference \cite{sheng2023flexgen,sun2025breaking,chen2025impress}. Nevertheless, the relatively low bandwidth of the SSD means that frequent data transfers between the SSD and host memory become the bottleneck in end-to-end latency.  

\noindent\textbf{KV Data Eviction for LLM Inference.} To address the memory and computation constraints in long-context inference, many prior works exploit the inherent sparsity of long-context inputs. During the decoding phase, these methods evaluate the importance of each token and selectively retain the KV data of important tokens on the GPU for subsequent inference, while evicting less important KV data to the CPU or dropping it entirely \cite{zhang2023h2o,zhao2024alisa,tang2024quest,lin2024infinite,sun2025breaking,gao2024cost,liu2023scissorhands,juravskyhydragen,ye2024chunkattention,zheng2024sglang}. However, directly applying these methods on AIPCs often faces fundamental challenges: the limited memory capacity may lead to out-of-memory (OOM) errors, while the low bandwidth of external storage makes frequent KV data transfers the dominant bottleneck to end-to-end inference latency. 

\noindent\textbf{KV Cache Compression and Quantization.} Another orthogonal line of research attempts to mitigate memory pressure by compressing the KV cache or quantizing it to lower precision (e.g., INT4 or INT2) \cite{liu2024kivi,li2024snapkv,ge2024model,liu2024minicache,caipyramidkv,liu2024cachegen,hooper2024kvquant,he2024zipcache}. While these software-based methods effectively reduce the peak memory footprint by a constant factor, they face fundamental limitations on memory-constrained AIPCs. First, aggressive quantization often sacrifices model accuracy and reasoning capability due to the loss of numerical precision and the truncation of activation outliers. Second, the KV cache fundamentally still grows linearly with the context length. For extremely long contexts (e.g., 32K or 128K), even a heavily compressed cache will inevitably exceed the available host and GPU memory limits of commodity PCs. Thus, software compression merely delays the memory wall rather than eliminating it.

\noindent\textbf{LLM Optimization using CSD.} Recently, several studies have focused on employing CSDs, e.g., SmartSSDs, to accelerate LLM-related workloads such as vector search \cite{tian2024scalable,liang2022vstore,kim2022accelerating,niu2024flashgnn}, data processing \cite{kang2013enabling,lee2020smartssd,soltaniyeh2022near,salamat2021nascent}, and inference \cite{pan2025instattention,deng2025kvnand,duan2025aegonkv}. Although prior work has demonstrated the benefits of CSDs, supporting long-context LLM inference on AIPCs remains challenging due to excessive ping-pong data movement and pronounced latency bottlenecks. Moreover, existing in-storage frameworks \cite{pan2025instattention,jang2025inf} offload exact attention computation, but resource-heavy Softmax operations quickly exhaust commodity FPGAs, necessitating custom hardware or multi-device offline batching. In contrast, HillInfer offloads only lightweight token evaluation, avoiding resource exhaustion to deliver real-time, long-context online serving on a single commodity SmartSSD.

\section{Conclusion}
In this paper, we presented HillInfer, a CSD-assisted KV eviction framework designed to break the memory and I/O walls of long-context LLM serving on memory-constrained AIPCs. By fundamentally rethinking in-storage processing, HillInfer decouples lightweight token evaluation from exact attention, successfully avoiding FPGA resource exhaustion on a single commodity SmartSSD. Our tightly integrated architecture—featuring the Hierarchical KV Cache Manager (HKM) to partition cache pools and eliminate I/O thrashing, the Adaptive Prefetch-based Pipeline (APP) to balance workloads and mask the straggler effect, and the CSD-based Evaluation Configuration (CEC) to enable resource-efficient near-data processing—achieves optimal I/O-compute overlap. Experimental evaluations confirm that HillInfer delivers up to an 8.56$\times$ speedup over existing baselines, providing low-latency, I/O-efficient long-context inference without compromising generation quality, thereby paving the way for scalable, privacy-preserving LLM deployment on AIPCs.





\bibliographystyle{plain}
\bibliography{sample}

\end{document}